%% file: main.tex
\documentclass[sigconf,natbib=true,screen=true]{acmart}

\acmSubmissionID{0849}

\input{Preamble/packages}
\input{Preamble/definitions}
\input{Preamble/copyright}
\input{Preamble/authors}

\input{Preamble/meta}

\begin{document}

\title[Ranked List Truncation for Large Language Model-based Re-Ranking]{Ranked List Truncation for \\ Large Language Model-based Re-Ranking}

\renewcommand{\shortauthors}{Chuan Meng, Negar Arabzadeh, Arian Askari, Mohammad Aliannejadi, and Maarten de Rijke}

\input{Sections/00_Abstract}

\maketitle

\acresetall

\input{Sections/01_Introduction}

\input{Sections/02_Motivation}

\input{Sections/03_Task_Definition}

\input{Sections/04_Method}
\input{Sections/05_Results}
\input{Sections/06_Related_Work}

\input{Sections/07_Conclusion}
\input{Sections/08_Acknowledgement}

\clearpage
\bibliographystyle{ACM-Reference-Format}
\balance
\bibliography{references}

%\clearpage
%\input{Sections/09_Backups}

\end{document}

%% file: Preamble/packages.tex
\usepackage{acronym}
\usepackage[skip=0pt]{caption}
\usepackage[inline]{enumitem}
\usepackage{booktabs} % For formal tables
\usepackage{multirow}
\usepackage{array}
\usepackage{pifont}
\usepackage{xspace}
\usepackage{subcaption}
\usepackage{makecell}
\usepackage{CJKutf8}

\usepackage[para]{footmisc}

%% file: Preamble/definitions.tex
\acrodef{DCG}{discounted cumulative gain}

\acrodef{EET}{efficiency-effectiveness trade-off}

\acrodef{RDE}{re-ranking depth estimation}

\acrodef{CEC}{certified error control}

\acrodef{VDP}{variable-depth pooling}
\acrodef{CS}{conversational search}

\acrodef{RLT}{ranked list truncation}

\acrodef{LtR}{learning-to-rank}

\acrodef{RAML}{reward augmented maximum likelihood}
\acrodef{LSTM}{long short-term memory}

\acrodef{LLM}{large language model}
\acrodef{QPP}{query performance prediction}
\acrodef{IR}{information retrieval}
\acrodef{NLP}{natural language processing}
\acrodef{PEFT}{parameter-efficient fine-tuning}
\acrodef{ICL}{in-context learning}
\acrodef{LoRA}{low-rank adaptation}

\acrodef{RR}{reciprocal rank}
\acrodef{AP}{Average Precision}
\acrodef{nDCG}{normalized discounted cumulative gain}

\acrodef{HSD}{Tukey's honestly significant difference}
\acrodef{ANOVA}{analysis of variance}

\acrodef{UEF}{utility estimation framework}
\acrodef{QPP-PRP}{pairwise rank preference-based QPP}
\acrodef{M-QPPF}{multi-task query performance prediction framework}
\acrodef{WRIG}{weighted relative information gain-based model}
\acrodef{NLP}{natural language processing}
\acrodef{CIS}{conversational information seeking}
\acrodef{ANCE}{Approximate nearest neighbor Negative Contrastive Estimation}

\acrodef{RL}{reinforcement learning}

\AtBeginDocument{%
  \providecommand\BibTeX{{%
    \normalfont B\kern-0.5em{\scshape i\kern-0.25em b}\kern-0.8em\TeX}}}

\newcommand{\moh}[1]{\textcolor{blue}{\textbf{[#1]}}}
\newcommand{\cm}[1]{\textcolor{red}{\textbf{[#1]}}}

\makeatletter
\@ifpackagelater{acronym}{2020/01/01}
{%
}
{%

\newcommand{\Ac}[1]{\ac{#1}(\textbf{properly capitalized})}
\newcommand{\Acf}[1]{\acf{#1}(\textbf{properly capitalized})}
}%
\makeatother

\newcommand{\header}[1]{\vspace*{.6mm}\noindent\textbf{#1}.}

\newcommand\setItemnumber[1]{\setcounter{enumi}{\numexpr#1-1\relax}}

\makeatletter
\g@addto@macro\normalsize{%
  \abovedisplayskip 1pt plus1pt %minus1pt%
  \belowdisplayskip 1pt plus1pt
  \abovedisplayshortskip  0pt plus1pt%
  \belowdisplayshortskip  0pt plus1pt% minus1pt%
}
\makeatother

%% file: Preamble/copyright.tex
\copyrightyear{2024}
\acmYear{2024}
\setcopyright{rightsretained}
\acmConference[SIGIR '24]{Proceedings of the 47th International ACM SIGIR Conference on Research and Development in Information Retrieval}{July 14--18, 2024}{Washington, DC, USA}
\acmBooktitle{Proceedings of the 47th International ACM SIGIR Conference on Research and Development in Information Retrieval (SIGIR '24), July 14--18, 2024, Washington, DC, USA}
\acmDOI{10.1145/3626772.3657864}
\acmISBN{979-8-4007-0431-4/24/07}

\makeatletter
\gdef\@copyrightpermission{
  \begin{minipage}{0.3\columnwidth}
   \href{https://creativecommons.org/licenses/by/4.0/}{\includegraphics[width=0.90\textwidth]{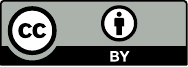}}
  \end{minipage}\hfill
  \begin{minipage}{0.7\columnwidth}
   \href{https://creativecommons.org/licenses/by/4.0/}{This work is licensed under a Creative Commons Attribution International 4.0 License.}
  \end{minipage}
  \vspace{5pt}
}
\makeatother

%% file: Preamble/authors.tex
\author{Chuan Meng}
\orcid{0000-0002-1434-7596}
\affiliation{%
  \institution{University of Amsterdam}
  \city{Amsterdam}
  \country{The Netherlands}
}
\email{c.meng@uva.nl}

\author{Negar Arabzadeh}
\orcid{0000-0002-4411-7089}
\affiliation{%
  \institution{University of Waterloo}
  \city{Waterloo}
  \country{Canada}
}
\email{narabzad@uwaterloo.ca}

\author{Arian Askari}
\orcid{0000-0003-4712-832X}
\affiliation{%
  \institution{Leiden University}
  \city{Leiden}
  \country{The Netherlands}
}
\email{a.askari@liacs.leidenuniv.nl}

\author{Mohammad Aliannejadi}
\orcid{0000-0002-9447-4172}
\affiliation{%
  \institution{University of Amsterdam}
  \city{Amsterdam}
  \country{The Netherlands}
}
\email{m.aliannejadi@uva.nl}

\author{Maarten de Rijke}
\orcid{0000-0002-1086-0202}
\affiliation{%
  \institution{University of Amsterdam}
  \city{Amsterdam}
  \country{The Netherlands}
}
\email{m.derijke@uva.nl}

%% file: Preamble/meta.tex
\ccsdesc[500]{Information systems~Retrieval models and ranking}
\ccsdesc[500]{Information systems~Language models}
\ccsdesc[500]{Information systems~Retrieval efficiency}
\ccsdesc[500]{Information systems~Retrieval effectiveness}
\keywords{Ranked list truncation, Re-ranking, Large language models}

%% file: Sections/00_Abstract.tex
\begin{abstract}
%\Ac{RLT} is a key task in \ac{IR}.
%The \ac{RLT} task aims to determine how many items in a ranked list to return, so as to optimize a user-defined metric. 
%
%Existing research has extensively studied \ac{RLT} for optimizing single-stage retrieval.
%
We study \ac{RLT} from a novel ``\textit{retrieve-then-re-rank}'' perspective, where we optimize re-ranking by truncating the retrieved list (i.e., trim re-ranking candidates).
\ac{RLT} is crucial for re-ranking as it can improve re-ranking efficiency by sending variable-length candidate lists to a re-ranker on a per-query basis. It also has the potential to improve re-ranking effectiveness.
Despite its importance, there is limited research into applying \ac{RLT} methods to this new perspective.
% \moh{two-stage ranking?}
%
To address this research gap, we reproduce existing \ac{RLT} methods in the context of re-ranking, especially newly emerged \ac{LLM}-based re-ranking.
In particular, we examine to what extent established findings on \ac{RLT} for retrieval are generalizable to the ``retrieve-then-re-rank'' setup from three perspectives:
\begin{enumerate*}[label=(\roman*)]
    \item assessing \ac{RLT} methods in the context of \ac{LLM}-based re-ranking with lexical first-stage retrieval,
    \item investigating the impact of different types of first-stage retrievers on \ac{RLT} methods, and 
    \item investigating the impact of different types of re-rankers on \ac{RLT} methods.
\end{enumerate*}
We perform experiments on the TREC 2019 and 2020 deep learning tracks, investigating 8 \ac{RLT} methods for pipelines involving 3 retrievers and 2 re-rankers.
%, leading to various configurations.
%
We reach new insights into \ac{RLT} methods in the context of re-ranking.
%and identify directions for future research.
%
%We open-source our supplementary materials at %\url{https://anonymous.4open.science/r/RLT4Reranking}.
\end{abstract}

%% file: Sections/01_Introduction.tex
\section{Introduction}
% ***Requirement from the SIGIR website***: The motivation for selecting the methods that are replicated or reproduced and the impact of these methods on the IR community
\Acf{RLT}, a.k.a.~ query cut-off prediction~\citep{cohen2021not,lesota2021modern}, has been studied for over two decades~\citep{arampatzis2009stop,manmatha2001modeling} and recently attracted lots of attention in the \ac{IR} community~\citep{bahri2023surprise,ma2022incorporating,wang2022mtcut,wu2021learning,bahri2020choppy,lien2019assumption}.
The task of \ac{RLT} is to determine how many items in a ranked list should be returned such that a user-defined metric is optimized~\citep{bahri2023surprise}.
%\ar{I think maybe you could prevent from saying "user-defined metric" by writing: The task of \ac{RLT} is to determine how many items in a ranked list should be returned, aiming to balance between the utility of search results and the cost of processing search results~\citep{bahri2023surprise}.}
The user-defined metric typically considers the balance between the utility of search results and the cost of processing search results~\citep{wang2022mtcut}.
\ac{RLT} is crucial in various \ac{IR} applications where it is money- or time-consuming to review a returned item~\citep{wu2021learning}.
E.g., in patent search~\citep{lupu2013patent} or legal search~\citep{wang2022mtcut,tomlinson2007overview}, providing a ranked list with an overwhelming number of items is too costly for patent experts or litigation support professionals~\citep{arampatzis2009stop}.

% ***Requirement from the SIGIR website***: The directions in which they try to generalize, chose different angles from the original work that they reproduce, 

\begin{figure*}[!t]
  \centering
  \includegraphics[width=1\linewidth]{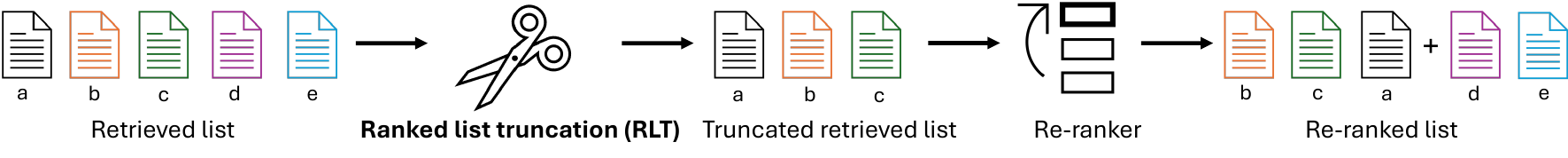}
  \caption{A schematic diagram of \ac{RLT} in the ``\textit{retrieve-then-re-rank}'' setup.}
    \label{fig:pipeline}
\end{figure*}

\header{A new angle}
Existing studies mainly focus on \ac{RLT} for single-stage retrieval, i.e., optimizing a user-defined metric (e.g., F1) of a retrieved list by truncating it at a certain position.
In this paper, we focus on \ac{RLT} for \textit{re-ranking}~\citep{askari2024self}, i.e., \ac{RLT} in a ``\textit{retrieve-then-re-rank}'' setup, as shown in Figure~\ref{fig:pipeline}. 
%However, there is limited research on \ac{RLT} for \textit{re-ranking}, i.e., \ac{RLT} in a ``\textit{retrieve-then-re-rank}'' setup.%~\citep{zamani2022stochastic,li2022certified,culpepper2016dynamic,wang2011cascade}.
%
In this setup, we still truncate a given retrieved list but focus on enhancing trade-offs between effectiveness and efficiency in re-ranking; truncating the retrieved list directly translates to a reduction in re-ranking depth. 
\ac{RLT} for re-ranking is important because: 
\begin{enumerate*}[label=(\roman*)]
    \item \ac{RLT} can improve re-ranking efficiency by sending variable-length lists of candidates to a re-ranker on a per-query basis; re-rankers are typically computationally expensive~\citep{lien2019assumption} and particularly, recently proposed \ac{LLM}-based re-rankers~\citep{ma2023fine,zhuang2023beyond,qin2023large,zhang2023rank,pradeep2023rankzephyr,pradeep2023rankvicuna,sun2023chatgpt} with billions of parameters lead to a substantial increase in computational overhead~\citep{zhuang2023setwise}, making it hard to apply them in practice; applying a fixed re-ranking cut-off to all queries is a common practice in the literature;
    however, individual queries can be answered effectively by a shorter or a longer list of re-ranking candidates~\citep{bruch2023efficient}, so \ac{RLT} can avoid unnecessary re-ranking costs by dynamically trimming the retrieved list;
    and
    \item \ac{RLT} has the potential to improve re-ranking effectiveness; indeed, feeding a long retrieved list that includes many irrelevant items to a re-ranker instead can result in inferior re-ranking quality than a shorter retrieved list~\citep{zamani2022stochastic} . 
    %% it's still too early to talk about our results; we're still in the stage of setting up the problem and goals:
    %<too early>
    %This finding is also consistent with our empirical analysis (see Section~\ref{}).
    % a deeper re-ranking depth does not always bring about any improvement, and even is detrimental for re-ranking performance, and
    %%% </too early>
\end{enumerate*} 

Despite its importance, limited research has explored the application of \ac{RLT} methods in the ``\textit{retrieve-then-re-rank}'' setup~\citep{zamani2022stochastic,li2022certified,culpepper2016dynamic,wang2011cascade}.
E.g., \citet{zamani2022stochastic} only use one \ac{RLT} method to truncate retrieved lists from BM25 to improve the performance of BERT-based re-ranking~\citep{nogueira2019passage}.
Put differently, there is a lack of systematic and comprehensive studies into the use of \ac{RLT} methods that have originally been introduced to optimize retrieval, in the context of re-ranking, especially newly emerged \acp{LLM}-based re-ranking.

\header{Research goal}
In this reproducibility paper, we examine \textit{to what extent established findings on \ac{RLT} for retrieval are generalizable to the ``retrieve-then-re-rank'' setup}.
Specifically, we study the following \textbf{findings} from the literature on \ac{RLT}:
\begin{enumerate*}[label=(\roman*)]
%\item truncating neural retriever performs better than truncating lexical retriever;
%\item document statistics information is beneficial~\citep{lien2019assumption}..
    \item Supervised \ac{RLT} methods generally perform better than their unsupervised counterparts (e.g., set a fixed cut-off for all queries)~\citep{wang2022mtcut,wu2021learning,bahri2020choppy,lien2019assumption}.

    \item Distribution-based supervised \ac{RLT} methods (i.e., directly predict a distribution among all candidate cut-off points) perform better than their sequential labeling-based counterpart (i.e., predict whether to truncate at each candidate point)~\citep{wang2022mtcut,wu2021learning,bahri2020choppy}.

    \item Jointly learning \ac{RLT} with other tasks (e.g., predicting the relevance of each item in the retrieved list) results in better \ac{RLT} quality~\citep{wang2022mtcut}. 

    \item When truncating a retrieved list returned by a neural-based retriever, incorporating its embeddings improves \ac{RLT} quality~\citep{ma2022incorporating}. 
\end{enumerate*}

% ***Requirement from the SIGIR website***: experimental setup(s) they select to support their research in these new directions;
\header{Reproducibility challenge}
We highlight the main challenges of applying \ac{RLT} methods from optimizing retrieval to optimizing re-ranking:
\begin{enumerate*}[label=(\roman*)]
    \item the new ``retrieve-then-re-rank'' setup leads to a new optimization goal for \ac{RLT} methods, i.e., improving the trade-offs between effectiveness and efficiency in the re-ranking process; more importantly, a specific trade-off can be considered as the optimization goal to meet the requirements of a specific scenario, e.g., effectiveness is more important than efficiency in professional search than web-search; and
    \item the re-ranking setup introduces the \emph{type} of re-ranker as a factor that influences \ac{RLT} quality; also, it is important to investigate the impact of the interaction between retrievers and re-rankers on \ac{RLT}; thus, it is important to explore \ac{RLT} performance under different pipelines of widely-used retrievers, e.g., lexical, leaned sparse~\citep{formal2022distillation} and dense~\citep{ma2023fine} retrievers, and different re-rankers, e.g., \ac{LLM}-based~\citep{ma2023fine} or pre-trained language model-based re-rankers~\citep{nogueira2020document}.
\end{enumerate*} 

\header{Scope}
We consider the challenges and examine each established finding from the literature on \ac{RLT} in three settings:
\begin{enumerate*}[label=(\roman*)]
\item we begin by checking if \ac{RLT} methods optimizing for different trade-offs between effectiveness and efficiency of a state-of-the-art \ac{LLM}-based re-ranker, RankLLaMA~\citep{ma2023fine}, with a lexical first-stage retriever; next, to study the impact of retriever types on \ac{RLT} methods, we assess \ac{RLT} methods for the \ac{LLM} re-ranker with other types of retrievers, i.e., learned sparse (SPLADE++~\citep{formal2022distillation}), and dense (RepLLaMA~\citep{ma2023fine}) retrievers; and
finally, to study the impact of the choice of re-rankers on \ac{RLT} methods, we assess \ac{RLT} methods for a widely-used pre-trained language model-based re-ranker, monoT5~\citep{nogueira2020document}.
%; given three types of retrievers, including lexical (BM25~\citep{robertson2009probabilistic}), learned sparse (SPLADE++~\citep{formal2022distillation}), and dense (RepLLaMA~\citep{ma2023fine}) retrievers, and a state-of-the-art \ac{LLM}-based re-ranker, RankLLaMA~\citep{ma2023fine}, 
%\item given identical pipelines, we assess \ac{RLT} methods optimized to model different trade-offs between effectiveness and efficiency in re-ranking; and 
%\item we finally check the impact of the re-ranker type on \ac{RLT} quality; given the identical retrievers, we assess those \ac{RLT} methods 
\end{enumerate*} 
We perform all experiments on the TREC 2019 and 2020 deep learning (TREC-DL) tracks~\citep{craswell2020,craswell2019} and consider 8 \ac{RLT} methods and pipelines involving 3 retrievers and 2 re-rankers, leading to various configurations.
%
%We aim to obtain new insights into existing \ac{RLT} methods in the ``\textit{retrieve-then-re-rank}'' setup, elaborate on their errors, and identify future directions. 

% The assumptions of the original work that they found to hold up, and the ones that could not be confirmed. The key is to share knowledge about what lessons from prior work held up
\header{Lessons}
Our experiments reveal that findings on \ac{RLT} do not generalize well to the ``\textit{retrieve-then-re-rank}'' setup.
E.g., we found supervised \ac{RLT} methods do not show a clear advantage over using a fixed re-ranking depth; potential fixed re-ranking depths are able to closely approximate the effectiveness/efficiency trade-offs achieved by supervised \ac{RLT} methods.
%distribution-based supervised methods generally perform better than their sequential labeling-based counterpart in most cases; 
%jointly learning \ac{RLT} with other tasks or incorporating neural retriever embeddings only helps in limited cases.
%
Moreover, we found the choice of retriever has a substantial impact on \ac{RLT} for re-ranking: with an effective retriever like SPLADE++ or RepLLaMA, a fixed re-ranking depth of 20 can already yield an excellent effectiveness/efficiency trade-off; increasing the fixed depth do not significantly improve effectiveness.
An error analysis reveals that supervised \ac{RLT} methods tend to fail to predict when not to carry out re-ranking; moreover, they seem to suffer from a lack of training data.
%The type of re-ranker does not appear to impact the findings.
%while one finding on \ac{RLT} for first-stage retrieval does not generalize to the ``\textit{retrieve-then-re-rank}'' setup well, two findings do generalize to the setup under some specific cases. 
%In short, unsupervised \ac{RLT} methods outperform supervised ones, and methods that optimize for \ac{RLT} jointly with other tasks can benefit from the joint training.

% Our findings are as follows:
% \begin{enumerate*}[label=(\roman*)]
% \item unsupervised \ac{RLT} methods generally perform better than supervised ones in terms of the joint consideration of re-ranking effectiveness and efficiency given pipelines of different types of retrievers and re-rankers;
% \item distribution-based supervised \ac{RLT} methods  perform better than their sequential labeling-based counterparts in most cases; 
% \item jointly learning \ac{RLT} with other tasks~\citep{wang2022mtcut} leads to a marked improvement in terms of re-ranking effectiveness/efficiency for the pipeline of SPLADE++--RankLLaMA, but does not benefit other combinations of other types of retrievers and re-rankers; and 
% \item the \ac{RLT} method incorporating results embeddings~\citep{ma2022incorporating} of a neural retriever performs well in re-ranking effectiveness when optimized for prioritizing effectiveness compared to other supervised \ac{RLT} methods, but does not improve efficiency markedly.
% \end{enumerate*} 

\header{Contributions}
Our main contributions are as follows:
\begin{itemize}[leftmargin=*,nosep]
%\item We identify a novel angle on \ac{RLT} in the \textit{``retrieve-then-re-rank''} setup.
\item We reproduce a comprehensive set of \ac{RLT} methods in a \textit{``retrieve-then-re-rank''} perspective.
\item We conduct an empirical analysis with a state-of-the-art \ac{LLM}-based re-ranker, revealing that setting fixed re-ranking cut-offs results in unnecessary computational costs and diminishes re-ranking quality.
\item We conduct extensive experiments on 2 datasets, 8 \ac{RLT} methods and pipelines involving 3 retrievers and 2 re-rankers, allowing a comprehensive understanding of how \ac{RLT} methods generalize to the new perspective.
We open source our code and data at \url{https://github.com/ChuanMeng/RLT4Reranking}.
\end{itemize}

%In this paper, we only consider open-source \ac{LLM}-based re-rankers because re-rankers based on commercial \acp{LLM}~\citep{sun2023chatgpt} (such as GPT-3.5/4) come with limitations like non-reproducibility, non-deterministic outputs, impeding their utility in academic settings~\citep{pradeep2023rankzephyr,zhang2023rank}
%it raises the concern that the current research findings are only applicable to the GPT models instead of the general LLMs.~\citep{zhang2023rank}

%--RankLLaMA pipeline
%SPLADE++--RankLLaMA pipeline
%RepLLaMA--RankLLaMA pipeline

%BM25--RankZephyr pipeline
%SPLADE++ ED--RankZephyr pipeline
%RepLLaMA--RankZephyr pipeline

%BM25--monoT5 pipeline
%SPLADE++ ED--monoT5 pipeline
%RepLLaMA--monoT5 pipeline

%\header{Method}
%We elect to leverage \acf{QPP} to solve the task of re-ranking depth estimation.
%\Ac{QPP} is an essential task in \acf{IR} and has long been studies for more than two decades~\citep{} in the \ac{IR} community.
%The task of \ac{QPP} aims to estimate the retrieval quality of a search system without human-generated relevance judgments.
%We introduce the task of re-ranking depth estimation for \ac{LLM}-based re-ranking, which aims to 

%% file: Sections/02_Motivation.tex
\begin{figure*}[!t]
    \centering
    \begin{subfigure}{0.5\columnwidth}
        \includegraphics[width=\linewidth]{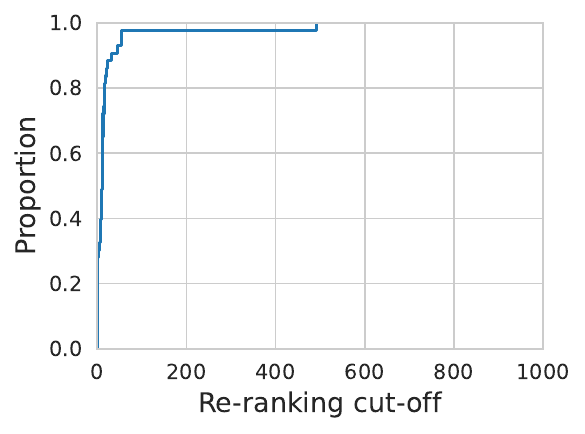}
        \vspace*{-7mm}
        \caption{TREC-DL 19}
        \label{fig:cdf-dl19-repllama-rankllama}
    \end{subfigure}
    \begin{subfigure}{0.5\columnwidth}
        \includegraphics[width=\linewidth]{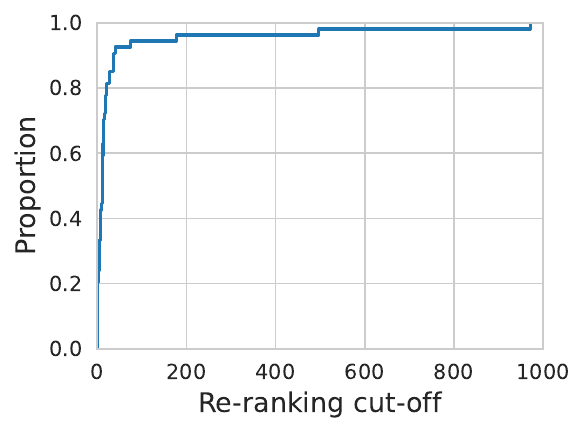}
        \vspace*{-7mm}
        \caption{TREC-DL 20}
        \label{fig:cdf-dl20-repllama-rankllama}
    \end{subfigure}
    \begin{subfigure}{0.5\columnwidth}
        \includegraphics[width=\linewidth]{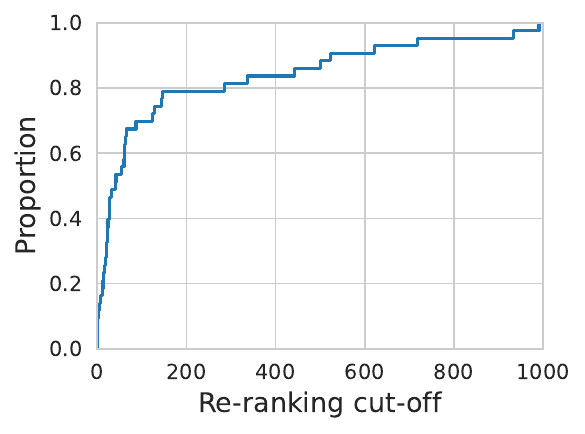}
        \vspace*{-7mm}
        \caption{TREC-DL 19}
        \label{fig:cdf-dl19-bm25-rankllama}
    \end{subfigure}
        \begin{subfigure}{0.5\columnwidth}
        \includegraphics[width=\linewidth]{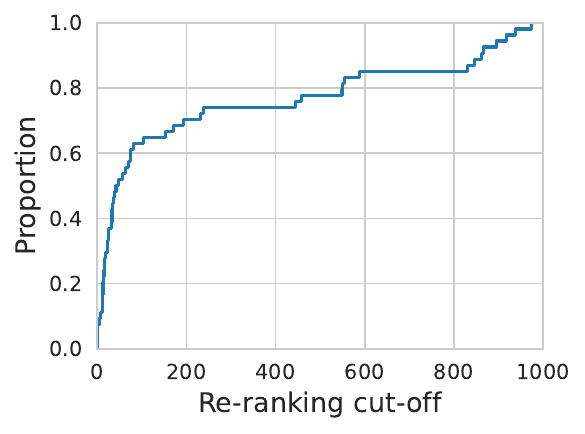}
        \vspace*{-7mm}
        \caption{TREC-DL 20}
        \label{fig:cdf-dl20-bm25-rankllama}
    \end{subfigure}
    \caption{
    Cumulative distribution function of oracle cut-offs for RepLLaMA--RankLLaMA (a, b) and BM25--RankLLaMA (c, d) on TREC-DL 19 and 20.
    The oracle cut-offs are the minimum re-ranking cut-offs that yield the highest nDCG@10 values.
    }
    \label{fig:cdf}
    \vspace*{1mm}
\end{figure*}

\begin{figure*}[!t]
    \centering
    \begin{subfigure}{0.5\columnwidth}
        \includegraphics[width=\linewidth]{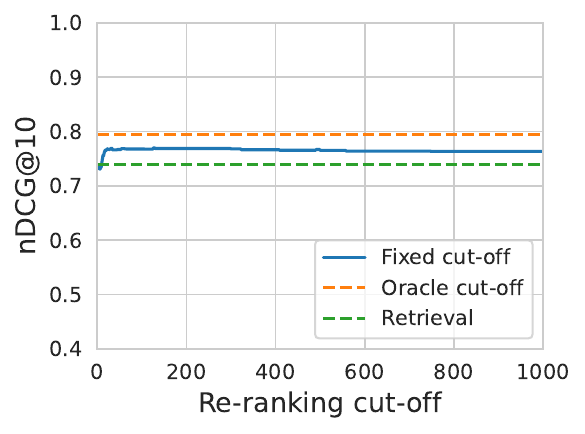}
        \vspace*{-7mm}
        \caption{TREC-DL 19}
        \label{fig:dl19-repllama-rankllama-ndcg}
    \end{subfigure}
    \begin{subfigure}{0.5\columnwidth}
        \includegraphics[width=\linewidth]{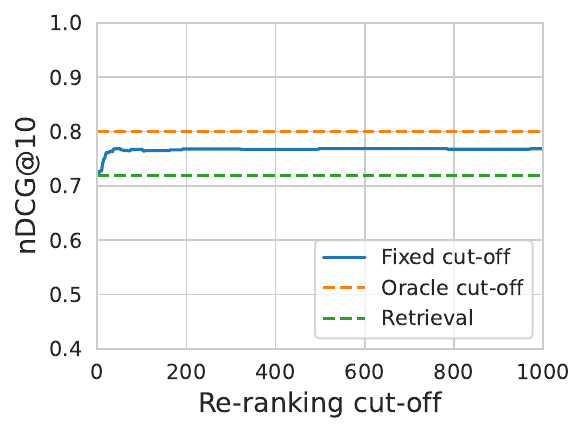}
        \vspace*{-7mm}
        \caption{TREC-DL 20}
        \label{fig:dl20-repllama-rankllama-ndcg}
    \end{subfigure}
    \begin{subfigure}{0.5\columnwidth}
        \includegraphics[width=\linewidth]{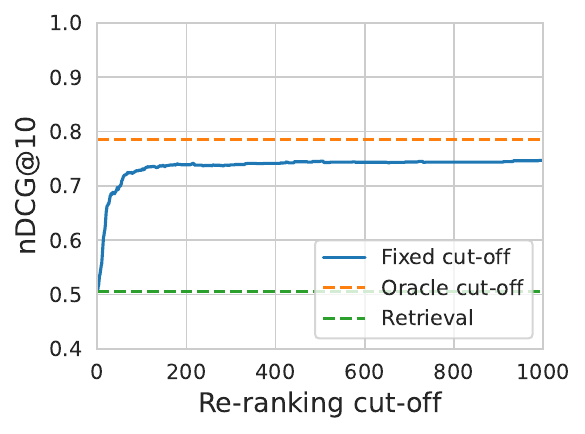}
        \vspace*{-7mm}
        \caption{TREC-DL 19}
        \label{fig:dl19-bm25-rankllama-ndcg}
    \end{subfigure}
    \begin{subfigure}{0.5\columnwidth}
        \includegraphics[width=\linewidth]{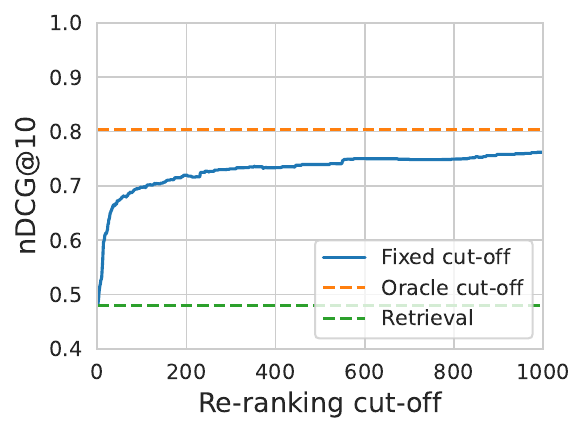}
        \vspace*{-7mm}
        \caption{TREC-DL 20}
        \label{fig:dl20-bm25-rankllama-ndcg}
    \end{subfigure}
    \caption{
    nDCG@10 values for RepLLaMA--RankLLaMA (a, b) and BM25--RankLLaMA (c, d) w.r.t.\ re-ranking cut-offs on TREC-DL 19 and 20.
    }
    \label{fig:ndcg}
    \vspace*{1mm}
\end{figure*}

\section{Motivation}
In the literature, applying a fixed re-ranking cut-off to all queries is a common practice~\citep{ma2023fine,zhuang2023beyond,qin2023large,zhang2023rank,pradeep2023rankzephyr,pradeep2023rankvicuna,ma2023zero,sun2023chatgpt}; however, individual queries may need a shorter or a longer list of re-ranking candidates~\citep{bruch2023efficient}.
%\moh{however, not all queries are equally hard and require the same re-ranking depth [REF]}.
%E.g.,  employ the \ac{LLM}-based re-ranker RankLLaMA to reorder the top 200 items returned by the \ac{LLM}-based retriever RepLLaMA.   
%We aim to describe the motivation for utilizing \ac{RLT} for re-ranking.
We conduct an empirical analysis to demonstrate how \ac{RLT} holds the potential to enhance both the effectiveness and efficiency in re-ranking compared to fixed cut-offs.
%
%To do so, we analyze two ``retrieve-then-re-rank'' pipelines on TREC-DL 19 and 20 datasets, namely, a state-of-the-art \ac{LLM}-based dense retriever (RepLLaMA)~\citep{ma2023fine} plus an \ac{LLM}-based re-ranker (RankLLaMA)~\citep{ma2023fine}, and BM25 plus RankLLaMA.
To do so, we analyze two ``retrieve-then-re-rank'' pipelines on the TREC-DL 19 and 20 datasets. We use an \ac{LLM}-based re-ranker (RankLLaMA~\citep{ma2023fine}) in both pipelines, but for first-stage retrieval, we employ a lexical retriever (BM25) in one pipeline and an \ac{LLM}-based dense retriever (RepLLaMA~\citep{ma2023fine}) in the other.

\header{Query-specific cut-offs improve efficiency} 
We study an \textit{oracle} setup in which we define the oracle as the minimum re-ranking cut-offs yielding the highest nDCG@10 values. 
we find that individual queries have different oracle cut-offs with a wide range. 
Thus, a fixed cut-off either wastes computational resources or compromises re-ranking quality for queries that need a deeper cut-off. 
Figure~\ref{fig:cdf} illustrates the cumulative distribution of oracle cut-offs for both pipelines on both datasets. 
Interestingly, about 30\% of queries do not need re-ranking with RepLLaMA as the retriever, and approximately 5\% with BM25; thus, calling expensive re-rankers can be omitted for these queries.
% ; we manually check those queries and find that in most cases the retrievers already achieve a perfect nDCG@10 value; thus, calling expensive re-rankers can be omitted.

\header{Query-specific cut-offs improve effectiveness} 
Figure~\ref{fig:ndcg} illustrates the comparison of re-ranking quality between using oracle and fixed cut-offs.
We find that oracle cut-offs always perform \textit{statistically significantly} (paired t-test, $p < 0.05$) better than all fixed cut-offs in terms of nDCG@10.
Hence, a deeper re-ranking cut-off does not consistently result in improvement and can even be detrimental to re-ranking quality.
Our finding is consistent with \citet{zamani2022stochastic}.
While one might argue that the re-ranking results with a deeper cut-off might be underestimated because of the limited number of judged items within the top 10 ranks, i.e., judged@10~\citep{pradeep2023rankzephyr}, we find that RankLLaMA's judged@10 values for using a fixed cut-off at 1000 and oracle cut-offs are similar, e.g., 95.35\% vs.96.05\% and 97.41\% vs. 97.41\% when RepLLaMA as the retriever on TREC-DL 19 and 20, respectively.

%adeeper re-ranking cut-off detrimentally impacts the results due to the limited number of the judged items within the top 10 ranks, 
%exhibits a high proportion using a fixed re-ranking cut-off at 1000: 95.35\% (95.81\%) and 97.41\% (98.15\%) when RepLLaMA (BM25) as the retriever on TREC-DL 19 and 20, respectively.

\ac{RLT} methods truncate the retrieved list (i.e., trim re-ranking candidates) on a per-query basis, suggesting that effective \ac{RLT} has the potential to improve re-ranking efficiency and effectiveness.

%% file: Sections/03_Task_Definition.tex
\section{Preliminaries and task definition}
We recap the task definition of \ac{RLT} and demonstrate the transition from optimizing retrieval to optimizing re-ranking.

\header{\ac{RLT} for retrieval}
Given a user query $q$, a collection $C$ containing $|C|$ items, and a retrieved list $L=[d_1, d_2, \ldots, d_{|L|}]$ with $|L|$ ($|L|\ll|C|$) items induced by a first-stage retriever over $C$ in response to $q$. 
An \ac{RLT} approach $f$ aims to predict a truncation point $k$ that maximizes a target metric that is about the retrieved list $L$ itself~\citep{ma2022incorporating,wang2022mtcut,wu2021learning,bahri2020choppy,lien2019assumption}, formally:
\begin{equation}
\begin{split}
k=f([x_1,x_2, \ldots, x_{|L|}])\in\{1,2,\ldots,|L|\},
\end{split}
\label{f1}
\end{equation}
where $[x_1,x_2, \ldots, x_{|L|}]$ are item features corresponding one-to-one with the items in the retrieved list $L=[d_1, d_2, \ldots, d_{|L|}]$. 
Typically, $x$ includes the retrieval score~\citep{bahri2020choppy} and item statistics~\citep{wang2022mtcut,wu2021learning,lien2019assumption}.
As for the target metric, $F1@k$ and $DCG@k$ have been widely used in prior studies~\citep{bahri2023surprise,ma2022incorporating,wang2022mtcut,wu2021learning,bahri2020choppy,lien2019assumption}.
E.g., $F1@k$ is calculated as:
\begin{equation}
\begin{split}
F1@k=& \frac{2\cdot P@k \cdot R@k}{P@k+R@k}, \\
P@k=&\frac{1}{k}\sum_{i=1}^{k}\mathbb{I}(y_i=1), R@K=\frac{1}{N_L}\sum_{i=1}^{k}\mathbb{I}(y_i=1), \\
\end{split}
\label{f1}
\end{equation}
where $y_i\in\{0,1\}$ is the relevance label for item $d_i$ in the truncated retrieved list, and $N_L$ denotes the number of relevant items in the retrieved list $L$.
%
\if0
Note that the original \ac{DCG} metric~\citep{jarvelin2002cumulated} is a monotonic metric since its value always increases with the value of $k$; it cannot evaluate \ac{RLT} properly because the optimal solution would be not to truncate at all~\citep{bahri2020choppy}.
Therefore, the \ac{DCG} metric employed in \ac{RLT} penalizes non-relevant items, rendering it a non-monotonic metric~\citep{bahri2023surprise,ma2022incorporating,wang2022mtcut,wu2021learning,bahri2020choppy}:
\begin{equation}
\begin{split}
DCG@k=\sum_{i=1}^{k} \frac{y_i}{\log_{2}(i+1)}~,
\end{split}
\label{dcg}
\end{equation}
where $y_i\in\{1,-1\}$; $y_i=-1$ if item $d_i$ is irrelevant to the query. 
\fi

\header{\ac{RLT} for re-ranking}
In the ``\textit{retrieve-then-re-rank}'' setup, we no longer focus on optimizing the retrieved list $L$, but we aim to test the capability of the \ac{RLT} in optimizing the trade-offs between effectiveness and efficiency in re-ranking.
As shown in Figure~\ref{fig:pipeline}, the truncated retrieved list $L_{1:k}=[d_1,\ldots,d_k]$, serving as re-ranking candidates, is further forwarded to a re-ranker that returns a re-ranked list $\hat{L}_{1:k}$.
We append the partial list \smash{$L_{k+1:|L|}$} from the retrieved list $L$ that is not re-ranked to \smash{$\hat{L}_{1:k}$}.

The target metric should evaluate the ranking quality of the re-ranked list $\hat{L}_{1:k}$ (with \smash{$L_{k+1:|L|}$}) in terms of an \ac{IR} evaluation metric (e.g., nDCG@10), and measure the computational cost of re-ranking.

%re-ranking depth estimation seeks to estimate the re-ranking depth $n \in [0, k]$ that leads to highest re-ranking performance.

%% file: Sections/04_Method.tex
\section{Reproducibility Methodology}
\label{model}

We state our research questions, the  experiments designed to address them, and our experimental setup.

%RQs
%1. dense relevance judgments vs sparse relevance judgments; different target metric?
%2. different first stage retrievers: quality candidate lists; score distribution
%3. the number of candidates to be truncated 1000 vs. 300
%4. balance between effectiveness and efficiency: more efficiency, more effectiveness

\subsection{Research questions and experimental design} 
%Our work is organized around the following research questions:
\begin{enumerate}[label=\textbf{RQ\arabic*},leftmargin=*]
    \setItemnumber{1}
    
    \item Do \ac{RLT} methods generalize to the context of \ac{LLM}-based re-ranking with a \textit{lexical first-stage retriever} when optimized for different effectiveness/efficiency trade-offs? \label{RQ1}
\end{enumerate}
To address \ref{RQ1}, we first quantify the trade-off between re-ranking effectiveness and efficiency, and then optimize \ac{RLT} methods to model different trade-offs between effectiveness and efficiency, simulating different requirements and scenarios; then, we evaluate their predicted truncation positions in terms of effectiveness and efficiency in \ac{LLM}-based re-ranking with a lexical retriever. \label{E1}
\begin{enumerate}[label=\textbf{RQ\arabic*},leftmargin=*]
    \setItemnumber{2}
    \item Do \ac{RLT} methods generalize to the context of \ac{LLM}-based re-ranking with \textit{learned sparse or dense first-stage} retrievers when optimized for the different trade-offs, and how does it compare to the case of a lexical retriever? \label{RQ2}
\end{enumerate}
For answering \ref{RQ2}, we still optimize \ac{RLT} methods for different trade-offs of the \ac{LLM}-based re-ranker used in \ref{RQ1}, but study their performance given learned sparse or dense retrievers, and compare the results with those of using a lexical retriever. \label{E2}
\begin{enumerate}[label=\textbf{RQ\arabic*},leftmargin=*]
    \setItemnumber{3}
    \item Do \ac{RLT} methods generalize to the context of \textit{pre-trained language model-based re-ranking}, and how does it compare to the case of an \ac{LLM}-based re-ranker? \label{RQ3}
\end{enumerate}
We address \ref{RQ3} by evaluating the truncation points predicted by \ac{RLT} methods w.r.t.\ effectiveness and efficiency in the context of a widely-used pre-trained language model-based re-ranker, and compare the results with those of the \ac{LLM}-based re-ranker. \label{E3}

\subsection{Experimental setup}
\label{sec:setup}

\header{\ac{RLT} approaches}
We reproduce a variety of unsupervised and supervised \ac{RLT} methods~\citep{bahri2023surprise,ma2022incorporating,wang2022mtcut,wu2021learning,bahri2020choppy,lien2019assumption}.
%
%Note that although candidate pruning methods~\citep{li2022certified,culpepper2016dynamic, wang2011cascade}, specifically designed for pruning the candidate list in a cascading ranking scheme, can be used in our scenario, previous studies study \ac{RLT} and candidate pruning separately and do not merge these two lines of research~\citep{bahri2023surprise,ma2022incorporating,wang2022mtcut,wu2021learning,bahri2020choppy,lien2019assumption}.
%
%In this work, our focus is on reproducing \ac{RLT} methods in ``\textit{retrieve-then-re-rank}'' setup.
%Exploring all approaches that have the potential to truncate the retrieved list and determining the optimal one is beyond the scope of our work.

%not on merging these two lines of research and we only focus on 

%As for unsupervised \ac{RLT} methods, we experiment with recent \ac{RLT} studies~\citep{bahri2023surprise,ma2022incorporating,wang2022mtcut,wu2021learning,bahri2020choppy}. 
%We do not consider assumption-based methods~\citep{arampatzis2009stop,manmatha2001modeling} as they model retrieval score distributions by fitting prior distributions, an assumption that does not always hold as retrieval settings change~\citep{wang2022mtcut,lien2019assumption}.
%
We consider the following unsupervised \ac{RLT} methods.
\begin{itemize}[leftmargin=*,nosep]
    \item \textbf{Fixed-$k$}~\citep{lien2019assumption} applies a fixed re-ranking cut-off to all queries; we follow common practice and consider cut-offs that are widely used in the literature about re-ranking, namely
    10~\citep{ma2023zero}, 20~\citep{pradeep2023rankzephyr,pradeep2023rankvicuna,ma2023zero}, 100~\citep{zhang2023rank,pradeep2023rankzephyr,pradeep2023rankvicuna,zhuang2023setwise,zhuang2023beyond,ma2023fine,zhuang2023open,drozdov2023parade,qin2023large,sun2023chatgpt}, 200~\citep{ma2023fine}, 1000~\citep{sachan2022improving}.
%
%\item Rank-based
%
%\item Score-based
%
%\item Mean-Max threshold
%
%\item \Ac{CEC}~\citep{li2022certified}
%
    \item \textbf{Greedy-$k$}~\citep{lien2019assumption} uses the fixed truncation position $k$ that maximizes the target metric value on a training set.
    \item \textbf{Surprise}~\citep{bahri2023surprise} first calibrates retrieval scores by using generalized Pareto distributions in extreme value theory~\citep{pickands1975statistical}, and truncates a ranked list using a score threshold.
\end{itemize} 

\noindent We consider the following supervised \ac{RLT} methods: 
\begin{itemize}[leftmargin=*,nosep]
    \item \textbf{BiCut}~\citep{lien2019assumption} is a \textit{sequential labeling-based} method; it uses a bidirectional \ac{LSTM} to encode item features over a ranked list, and optimizes the \ac{LSTM} make a binary prediction (continue or truncate) at each position in a ranked list.
%with a loss function that serves as a proxy to a target metric (e.g., F1)
%
    \item \textbf{Choppy}~\citep{bahri2020choppy} is a \textit{distribution-based} method, which directly predicts the distribution among all candidate cut-off points, using a transformer encoder~\citep{vaswani2017attention} to encode item features over a ranked list and predicts the distribution.
%; they train the model by a loss function that optimizes the expected value of a target metric. 
%
    \item \textbf{AttnCut}~\citep{wu2021learning} is also \textit{distribution-based}, encoding item features using a bidirectional \ac{LSTM} and a transformer encoder. \looseness=-1
%improve the optimization by employing \ac{RAML} to optimize a target metric directly. 
%
    \item \textbf{MtCut}~\citep{wang2022mtcut} is also \textit{distribution-based} and similar to AttnCut, but jointly trains the \ac{RLT} task with two auxiliary tasks: predicting the relevance of each item in the ranked list and increasing the margin between relevant and irrelevant items. We use the MMoECut variant due to its superior performance.
    \item \textbf{LeCut}~\citep{ma2022incorporating} is another \textit{distribution-based} and similar to AttnCut, but can only work with a neural-based retriever and incorporates its query--item embeddings as one of the item features. \citet{ma2022incorporating} further optimize LeCut with an \ac{LtR} model jointly. We omit this phase for a fair comparison since other methods are trained without signals from an external \ac{LtR} model.
\end{itemize}  

\noindent We also include \textbf{Oracle}, which truncates the retrieved list at the earliest position maximizing re-ranking quality per query.

\header{Optimizing effectiveness/efficiency trade-offs}
The leading challenge of adapting \ac{RLT} methods in the context of re-ranking is to optimize \ac{RLT} methods with a specific trade-off between effectiveness and efficiency.
To solve this challenge, we need to score each truncation point (i.e., re-ranking candidate cut-off) under different effectiveness/efficiency trade-offs.
%efine \ac{RLT} optimization objectives under different trade-offs; the optimization objectives should contain scores across all re-ranking cut-off candidates under different effectiveness/efficiency trade-offs
%
%specifically, each re-ranking cut-off candidate needs a different score under different effectiveness/efficiency trade-offs.
%
To do so, we quantify different trade-offs using the \ac{EET} metric~\citep{wang2010learning} and then compute \ac{EET} scores at a specific trade-off for all re-ranking candidate cut-offs.

\ac{EET} is defined for as the weighted harmonic mean of  effectiveness $\sigma$ and efficiency $\gamma$ measures: 
\begin{equation}
\begin{split}
EET= \frac{(1+\beta^2)\cdot(\gamma\cdot\sigma)}{\beta^2\cdot\sigma+\gamma},
\end{split}
\label{eet}
\end{equation}
where $\beta$ is a hyperparameter to control the relative importance of effectiveness and efficiency, where $\beta=0$ only considers effectiveness and as it increases more attention is paid to to efficiency.
\ac{EET} requires instantiation of $\sigma$ and $\gamma$ based on the specific use case~\citep{wang2010learning}.
We follow~\citep{wang2010learning} to instantiate efficiency $\gamma$ using ``exponential decay'':
\begin{equation}
\begin{split}
\gamma=\exp(\alpha\cdot k), 
\end{split}
\label{efficiency} 
\end{equation}
where $k\in\{1,2,\ldots,|L|\}$ is a truncation point (i.e., re-ranking candidate cut-off) in the given retrieved list $L$, and $\alpha\textless0$ is a hyperparameter to control how rapidly the efficiency measure decreases; we set $\alpha$ to -0.001.
We instantiate effectiveness $\sigma$ as the re-ranking gain with a cut-off $k$, which is quantified by the difference of re-ranking performance with a cut-off $k$ minus the performance without re-ranking; the performance is in terms of an \ac{IR} evaluation metric (e.g., nDCG@10).

\begin{figure}[!t]
    \centering
    \begin{subfigure}{0.495\columnwidth}
        \includegraphics[width=\linewidth]{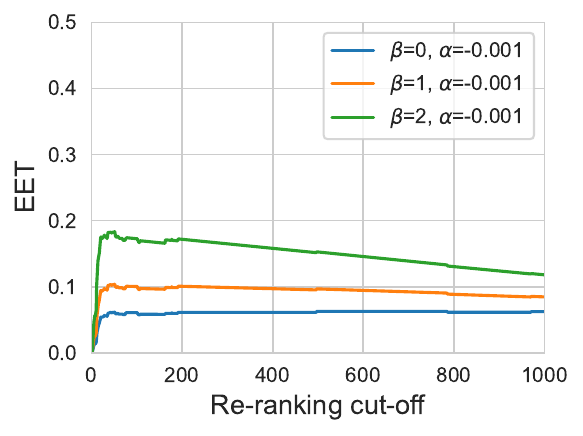}
        \vspace*{-7mm}
        \caption{RepLLaMA--RankLLaMA}
        \label{fig:eet-dl19-repllama-rankllama}
    \end{subfigure}
    \begin{subfigure}{0.495\columnwidth}
        \includegraphics[width=\linewidth]{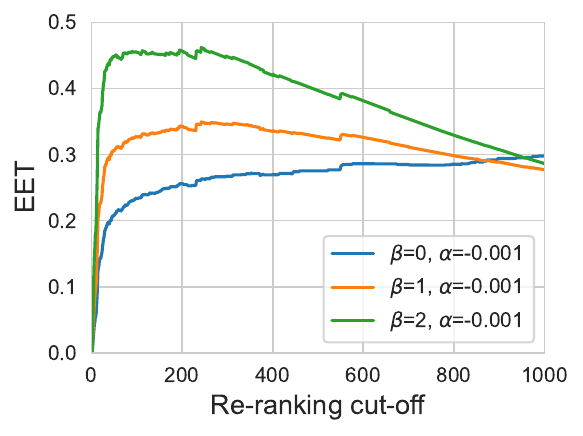}
        \vspace*{-7mm}
        \caption{BM25--RankLLaMA}
        \label{fig:eet-dl19-bm25-rankllama}
    \end{subfigure}
    \caption{
    Average \ac{EET} values across TREC-DL 20 queries w.r.t.\ re-ranking cut-offs. We use nDCG@10 in effectiveness $\sigma$. $\beta$ values 0, 1 and 2 represent prioritizing effectiveness, balancing effectiveness and efficiency, and emphasizing efficiency, respectively.
    }
    \label{fig:eet}
    \vspace*{1mm}
\end{figure}

Therefore, we can adjust $\beta$ in Equation~\ref{eet} and $\alpha$ in Equation~\ref{efficiency} in \ac{EET} to quantify different effectiveness/efficiency trade-offs in re-ranking, so as to generate \ac{EET} value distributions across all cut-off candidates under different trade-offs.
As illustrated in Figure~\ref{fig:eet}, we consider three trade-offs between effectiveness and efficiency: $\beta$=0 (emphasizing effectiveness), 1~(weighting effectiveness and efficiency equally), and 2 (prioritizing efficiency).
With the help of \ac{EET} value distributions under the three trade-offs, we optimize all distribution-based \ac{RLT} methods (Choppy~\citep{bahri2020choppy}, AttnCut~\citep{wu2021learning}, MtCut~\citep{wang2022mtcut}, LeCut~\citep{ma2022incorporating}) and Greedy-$k$ for the three trade-offs.

However, the sequential labeling-based \ac{RLT} method BiCut~\citep{lien2019assumption} cannot optimize a \ac{EET} value distribution.
During training, BiCut optimizes the probability of ``continue'' and ``truncation'' at each position in a ranked list via the following loss:
\begin{equation}
\begin{split}
\mathcal{L}=\sum_{i=1}^{|L|}(\eta\mathbb{I}(y_i=0)\frac{p_i}{1-r}+(1-\eta)\frac{1-p_i}{r}\mathbb{I}(y_i=1)),
\end{split}
\label{bicut_loss} 
\end{equation}
where $y_i\in\{0,1\}$ is the relevance label for an item at a position $it$, h $p_i$ is the ``continue'' probability at a position $i$, and $r$ is the proportion of relevant items in the ranked list; $\eta\in[0,1]$ is a hyperparameter to control the balance between ``continue'' and ``truncation''.
We optimize BiCut for different effectiveness/efficiency trade-offs by adjusting $\eta$ values, e.g., BiCut trained with a high $\eta$ value tends to truncate a retrieved list earlier, resulting in more efficiency. 
Specifically, we consider $\eta$=0.4 (emphasizing effectiveness), 0.5~(balancing effectiveness and efficiency), and 0.6 (prioritizing efficiency).\footnote{
We also explore $\eta$ values of 0.3 and 0.7. BiCut trained with the former tends not to truncate the retrieved list at all, while BiCut trained with the latter tends to truncate the entire retrieved list.}

\header{Datasets}
We experiment with 2 widely-used \ac{IR} datasets, TREC 2019 and 2020 deep learning (TREC-DL) tracks~\citep{craswell2020,craswell2019}.\footnote{We also conducted experiments on Robust04 and draw a similar conclusion as TREC-DL 19 and 20; due to space constraints, we show the result on Robust04 in our repository.} 
These datasets offer relevance judgments in multi-graded relevance scales per query.
TREC-DL 19 and 20 are built upon MS MARCO V1 passage ranking collection encompassing 8.8 million passages.

\header{Choice of retrievers}
% for \ref{RQ1}, \ref{RQ2}, and \ref{RQ3}
Regarding retrievers, we employ three distinct types: a lexical-based retriever BM25~\citep{robertson2009probabilistic}, a learned sparse retriever SPLADE++~(``EnsembleDistil'')~\citep{formal2022distillation} and an \ac{LLM}-based dense retriever RepLLaMA (7B)~\citep{ma2023fine}.
To increase the comparability and reproducibility of our paper, we obtain retrieval results of BM25 and SPLADE++ using the publicly available resource from Pyserini\footnote{\url{https://github.com/castorini/pyserini}} and get retrieval results of RepLLaMA from Tevatron;\footnote{\url{https://github.com/texttron/tevatron}} each retriever returns 1000 items per query.

\header{Choice of re-rankers}
For \ref{RQ1}, \ref{RQ2}, we employ a state-of-the-art \ac{LLM}-based point-wise reranker, RankLLaMA (7B)~\citep{ma2023fine} and use the resource from Tevatron.
For \ref{RQ3}, we employ a widely-used pre-trained language model-based re-ranker, monoT5 (``monot5-base-msmarco'')~\citep{nogueira2020document} and use the resource from PyGaggle.\footnote{\url{https://github.com/castorini/pygaggle}}

%and consider two first-stage retrievers and two \ac{LLM}-based re-rankers.
%For retrievers, we consider a lexical-based one, BM25, and an \ac{LLM}-based one, RepLLaMA~\citep{ma2023fine}, which achieves state-or-the-art retrieval performance.
%For \ac{LLM}-based re-rankers, we only consider point-wise re-rankers.
%Specifically, we consider a \textit{special token-based} one, RankLLaMA~\citep{ma2023fine} and a \textit{query likelihood-based} one.

\header{Evaluation metrics}
We measure re-ranking effectiveness using nDCG@10, the official evaluation metric in TREC deep learning tracks~\citep{craswell2020, craswell2019}, and a widely employed metric in ranking literature~\citep{pradeep2023rankzephyr,ma2023fine,nogueira2020document}.
We follow \citep{zhuang2023setwise} to evaluate re-ranking efficiency by calculating the average re-ranking cut-off across all test set queries, i.e., the number of the average number of re-ranking inferences per query.
This consideration is driven by the fact that the re-rankers we employ in this paper are point-wise, and the time spent in point-wise re-ranking is directly proportional to the length of a re-ranking cut-off (i.e., the length of a truncated retrieved list)~\citep{macavaney2022adaptive}.
We additionally gauge the efficiency by measuring per-query latency; all latency measurements exclude the time to load data and models.
We do not consider the latency of first-stage retrieval.
Note that \ac{RLT} methods are lightweight with significantly fewer parameters compared to state-of-the-art re-rankers; the latency introduced by \ac{RLT} methods can be neglected in the "retrieve-then-re-rank" setup.
%
%All latency measurements exclude the time to load data and models.
%Component results are summed to yield end-to-end query latency
%We report the per-query latency for each configuration and compute a speedup by normalizing against the latency of each baseline.

%To capture variance, we repeated runs five times.

% we did not use a distributed retrieval setting over multiple smaller indices coordinated by a broker (as might be deployed in a real-world system) as we did not wish to consider unrelated issues such as network latencies

\header{Implementation details}
%To implement \ac{RLT} methods in the context of re-ranking, we refer to the code\footnote{\url{https://github.com/myx666/LeCut}} for \ac{RLT} in single-stage retrieval, released by \citet{ma2022incorporating}.
We pass the top-1000 retrieved items to all \ac{RLT} methods per query because 1000 is typically the deepest re-ranking depth in the literature~\citep{ma2023fine,sachan2022improving,nogueira2019multi}.
Note that Suprise~\citep{bahri2023surprise} only depends on retrieval scores and uses a score threshold for truncation; the score threshold, set based on Cramer-von-Mises statistic testings~\citep{darling1957kolmogorov}, is not a tunable hyperparameter; thus, Suprise cannot be tuned for different effectiveness/efficiency trade-offs. 

%outputs a cut-off threshold for relevance score on a per-query basis. It starts with the smallest non-relevant score, and the method iteratively moves upwards through the scores to identify the optimal cutoff point. 
%We note that this method is only dependent on the list of relevance scores and thus unlike the supervised methods cannot be tuned for different trade-offs. The cut-off threshold in Surprise baseline would be set based on Cramer-von-Mises statistic testings \cite{darling1957kolmogorov} and thus it is indepenedent of external hyperparameters.

For all supervised \ac{RLT} methods, we use identical item features to eliminate confounding factors from the input; each item is represented by its retrieval score, length, unique token count, and the cosine similarity between its tf-idf/doc2vec~\citep{le2014distributed} vector and the vectors of its adjacent items.
We follow \citep{wu2021learning,wang2022mtcut} to use gensim\footnote{\url{https://radimrehurek.com/gensim}} for computing tf-idf and doc2vec vectors for each item; The dimension of the tf-idf vectors on the MS MARCO V1 collection is 846,221; we follow \citep{wang2022mtcut} to set the dimension of doc2vec vectors as 128.
LeCut relies on query-item embeddings from a neural retriever as extra item features; thus, we provide LeCut with the concatenation of query and item embeddings from RepLLaMA.
Note that we follow all original work to set hyperparameters:
for BiCut, we set \# Bi-LSTM layers to 2 and the LSTM hidden size to 128; we train BiCut via Adam~\citep{kingma2014adam} with a learning rate of $1\times10^{-4}$;
for Choppy, we set \# transformer layers to 3, \# transformer heads to 8, and transformer hidden size to 128; we train Choppy via Adam with a learning rate of $1\times10^{-3}$;
for AttnCut, we set \# Bi-LSTM layers to 2, the LSTM hidden size to 128, \# transformer heads to 4, and the transformer hidden size to 128; MtCut and LeCut share almost the same hyperparameters with AttnCut; we train AttnCut, MtCut and LeCut via Adam with a learning rate of $3\times10^{-5}$.
We train all supervised \ac{RLT} methods for 100 epochs using a batch size of 64 on TREC-DL 19, then infer them on TREC-DL 20, and vice versa.
All methods are trained/inferred on an NVIDIA A100 GPU (40GB).

%% file: Sections/05_Results.tex
\section{Results and discussions}
\label{res}

\if0
\begin{figure*}[!t]
  \centering
  \includegraphics[width=.9\linewidth]{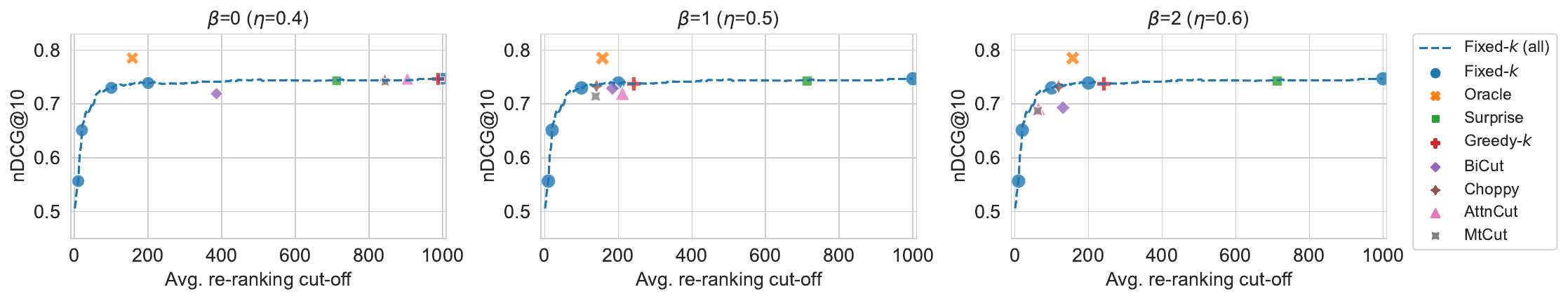}
  \caption{BM25--RankLLaMA on TREC-DL 19.}
    \label{fig:bm25_rankllama_dl19}
\end{figure*}
\fi

\begin{figure*}[!t]
  \centering
  \includegraphics[width=1\linewidth]{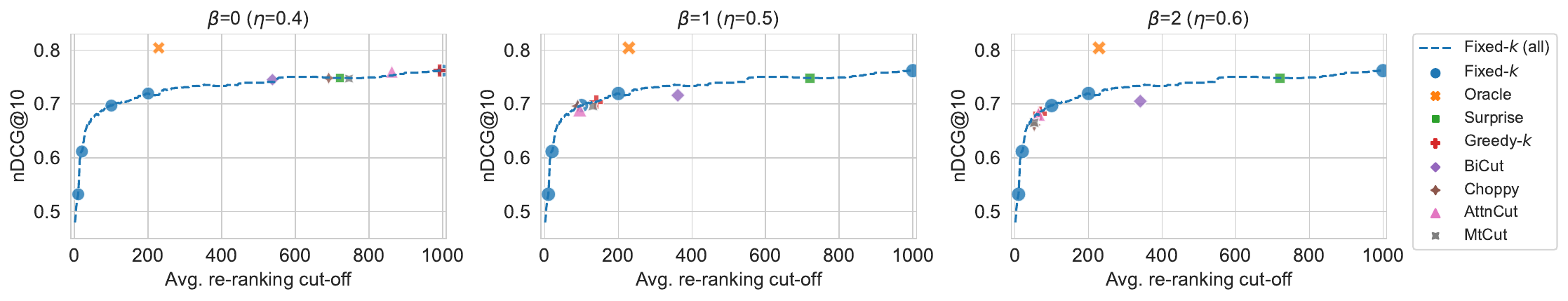}
  \caption{A comparison of RLT methods in predicting re-ranking cut-off points for BM25--RankLLaMA on TREC-DL 20. 
  $\beta$=0 ($\eta$=0.4), $\beta$=1 ($\eta$=0.5), and $\beta$=2 ($\eta$=0.6) represent effectiveness emphasis, balance, and efficiency emphasis, respectively.}
    \label{fig:bm25_rankllama_dl20}
\end{figure*}

\begin{table}[!t]
\centering
\caption{
A comparison of \ac{RLT} methods in predicting re-ranking cut-off points for BM25--RankLLaMA on TREC-DL 19 and 20. 
Query latency is measured in seconds.
The best nDCG@10 value in each column is marked in \textbf{bold}, and the second best is \underline{underlined}.
Significant differences with Fixed-$k$ (100), Fixed-$k$ (200) and Fixed-$k$ (1000) are marked with $^*$, $^\S$ and $^\dagger$, respectively (paired t-test, $p < 0.05$).
}
\label{tab:bm25_rankllama}
\setlength{\tabcolsep}{.6mm}
\resizebox{\columnwidth}{!}{
\begin{tabular}{l @{~} rcr rcr} 
\toprule
 \multirow{2}{*}{Method} & \multicolumn{3}{c}{TREC-DL 19} & \multicolumn{3}{c}{TREC-DL 20}  \\
  \cmidrule(lr){2-4} \cmidrule(lr){5-7}
& \multicolumn{1}{c}{Avg. k} & \multicolumn{1}{c}{nDCG@10}     & \multicolumn{1}{c}{Lat.}  & \multicolumn{1}{c}{Avg. k}   & \multicolumn{1}{c}{nDCG@10}    & \multicolumn{1}{c}{Lat.}     \\
\midrule
w/o re-ranking & -  & \phantom{000}0.506$^*$$^\S$$^\dagger$    & -   & -   & \phantom{000}0.480$^*$$^\S$$^\dagger$  &  -  \\
  \midrule
  Fixed-$k$ (10)  & 10   & \phantom{000}0.557$^*$$^\S$$^\dagger$    & 0.30 & 10  &  \phantom{000}0.532$^*$$^\S$$^\dagger$  &  0.30    \\
  Fixed-$k$ (20) & 20   & \phantom{000}0.651$^*$$^\S$$^\dagger$    & 0.60  &  20 & \phantom{000}0.612$^*$$^\S$$^\dagger$   & 0.60  \\
  Fixed-$k$ (100) & 100  & 0.730    & 2.98  & 100  &\phantom{00}0.697$^\S$$^\dagger$  & 2.98   \\
  Fixed-$k$ (200) & 200  & 0.739   & 5.95  & 200  & \phantom{00}0.719$^*$$^\dagger$  & 5.96   \\
 %$k=300$ & 300  & 0.739   &  & 300 & 0.731  &    \\
 %$k=400$ & 400  & 0.741    &  & 400  & 0.733  &  \
 %$k=500$ & 500   & 0.745   &  & 500 & 0.739  &   \\
% $k=600$ & 600  &  0.743    &  & 600 & 0.750  &   \\
 Fixed-$k$ (1000)  & 1000 & \textbf{0.747}    & 29.77  & 1000  &  \phantom{00}\textbf{0.762}$^*$$^\S$ & 29.78 \\
Surprise & 712  &  0.743    & 21.20 & 721   &  \phantom{000}0.748$^*$$^\S$$^\dagger$     & 21.46  \\
\midrule
Greedy-$k$ ($\beta$=0)  & 987   &  0.746     & 29.39 & 992   & \phantom{00}\textbf{0.762}$^*$$^\S$      & 29.54     \\
BiCut ($\eta$=0.4)  & 386     & 0.719     & 11.48     & 538  & \phantom{0}0.745$^\dagger$  &  16.01 \\
Choppy ($\beta$=0)  & 843     & \underline{0.744}   & 25.10  &  690  & \phantom{00}0.748$^*$$^\S$      & 20.56  \\
AttnCut ($\beta$=0) &  904    &  \textbf{0.747}   & 26.92 & 862  & \phantom{00}\underline{0.760}$^*$$^\S$      & 25.67   \\
MtCut ($\beta$=0)  & 844    & 0.741    & 25.12 & 745   & \phantom{000}0.747$^*$$^\S$$^\dagger$      & 22.20  \\
\midrule
%$\alpha$=-0.001 &  &  &  & & & \\
Greedy-$k$ ($\beta$=1) & 242     & 0.737   & 7.21  & 140  &  \phantom{00}0.703$^\S$$^\dagger$     &  4.17    \\
BiCut ($\eta$=0.5)  & 184     &  \phantom{0}0.729$^\dagger$    &  5.48    & 362  & \phantom{0}0.716$^\dagger$     &  10.77 \\
Choppy ($\beta$=1) & 141     &  0.733     & 4.19 & 90    & \phantom{00}0.696$^\S$$^\dagger$       & 2.67 \\
AttnCut ($\beta$=1) &  211    & \phantom{00}0.720$^\S$$^\dagger$      & 6.29 & 95    & \phantom{00}0.689$^\S$$^\dagger$      & 2.83  \\
MtCut ($\beta$=1) & 138     &  \phantom{00}0.714$^\S$$^\dagger$     & 4.12 & 131   & \phantom{0}0.696$^\dagger$      &  3.90\\
%$\alpha$=-0.01 &  &  &  & & & \\
%Greedy-$k$ ($\beta$=1) & 45     & 0.697      &  & 61   &  0.681     &      \\
%Greedy-$k$ ($\beta$=2)  & 29     &  0.681     &  &  29  &  0.650     &      \\
%\midrule
%BiCut ($\alpha$=0.45)  & 168.40     &  0.700    &      & 501.56  & 0.741     &   \\

%BiCut ($\alpha$=0.55)  & 130.49     & 0.688     &      & 412.59  & 0.723     &   \\

%BiCut ($\alpha$=0.65)  & 99.09     &  0.684    &      & 204.69  & 0.692     &   \\
%BiCut ($\alpha$=0.70)  & 34.67     & 0.596     &      & 124.74  & 0.650      &   \\
%BiCut ($\alpha$=0.75)  & 10.28     & 0.539     &      & 41.74  & 0.533     &  \\
%BiCut ($\alpha$=0.80) & 0.00     &  0.506    &      & 0.0  &  0.48    &   \\
\midrule
Greedy-$k$ ($\beta$=2)  & 242     & 0.737      & 7.21 &  68  & \phantom{00}0.682$^\S$$^\dagger$      & 2.03      \\
BiCut ($\eta$=0.6)  & 131     &  \phantom{000}0.693$^*$$^\S$$^\dagger$    &  3.89    & 341 & \phantom{0}0.705$^\dagger$     & 10.15  \\
Choppy ($\beta$=2)  & 119     & 0.732      & 3.54 & 53   &  \phantom{000}0.661$^*$$^\S$$^\dagger$     & 1.57  \\
AttnCut ($\beta$=2)  &  64     &  \phantom{000}0.692$^*$$^\S$$^\dagger$     & 1.91 & 64  & \phantom{00}0.681$^\S$$^\dagger$     &  1.90  \\
MtCut ($\beta$=2) &  62    & \phantom{000}0.687$^*$$^\dagger$$^\S$      & 1.85 & 52   &  \phantom{00}0.665$^\S$$^\dagger$     & 1.56 \\
\midrule
%$\alpha$=-0.001 &  &  &  & & & \\
%\alpha$=-0.01 &  &  &  & & & \\
%Choppy ($\beta$=1)  &  43.53   &  0.694    &  &  45.44  &  0.666     &  \\
%AttnCut ($\beta$=1) & 26.58     &  0.653     &  & 35.06   &  0.639    &   \\
%MtCut ($\beta$=1)  & 41.30     & 0.613      &  & 30.09   & 0.615      &  \\
%Choppy ($\beta$=2)  &  27.67    & 0.670      &  & 29.17   &  0.627     &  \\
%AttnCut ($\beta$=2)  & 38.35     & 0.617      &  & 24.24   & 0.625      &   \\
%MtCut ($\beta$=2)  & 36.58     & 0.593      &  & 18.09   & 0.595     &  \\
%\midrule
%$\alpha$=0.05 &  &  &  & & & \\
%Choppy ($\beta$=1)  & 14.0   & 0.605    &  & 21.26   &  0.613     &  \\
%AttnCut ($\beta$=1) & 9.44     & 0.545      &  &  9.87  & 0.531      &   \\
%MtCut ($\beta$=1)  & 32.70     &  0.565     &  &  11.61  &  0.559     &  \\
%Choppy ($\beta$=2)  & 14.0     &  0.605   &  & 16.52   & 0.603       &  \\
%AttnCut ($\beta$=2)  & 8.51   & 0.544      &  & 8.65  & 0.519      &   \\
%MtCut ($\beta$=2)  & 8.51     &  0.555      &  &  7.57  & 0.536      &  \\
%\midrule
  Oracle & 157  &  \phantom{000}0.785$^*$$^\S$$^\dagger$  & 4.67  & 229  & \phantom{000}0.804$^*$$^\S$$^\dagger$  & 6.80 \\
\bottomrule
\end{tabular}
}
\end{table}

\subsection{\ac{RLT} for \ac{LLM}-based re-ranking}
\label{res:rq1}

To answer \ref{RQ1}, we evaluate \ac{RLT} methods (optimized for the three effectiveness/efficiency trade-offs) in the context of an \ac{LLM}-based re-ranker (RankLLaMA~\citep{ma2023fine}) with a lexical retriever (BM25).
We report the results on TREC-DL 19 and 20 in Table~\ref{tab:bm25_rankllama}; we also plot the result on TREC-DL 20 in Figure~\ref{fig:bm25_rankllama_dl20}.
%\footnote{Note that we omit the significance test due to the lack of a metric for assessing \ac{RLT} in tackling the trade-off between efficiency and effectiveness. 
%We believe there is a need to design such a metric in future work to quantify this balance, enabling statistically significant t-test comparisons between methods.}
We have three observations.

First, compared to unsupervised \ac{RLT} methods, supervised \ac{RLT} one only shows an advantage at achieving better re-ranking effectiveness at a less re-ranking cost in the scenario emphasizing effectiveness; nevertheless, alternative fixed re-ranking depths can deliver results on par with those obtained through supervised methods.
\begin{enumerate*}[label=(\roman*)]
    \item In the scenario where effectiveness is prioritized ($\beta$=0 and $\eta$=0.4), supervised \ac{RLT} methods achieve better re-ranking effectiveness while maintaining less re-ranking cost.
    For instance, Choppy ($\beta$=0) and AttnCut ($\beta$=0) show no significant difference from Fixed-$k$ (1000) in terms of nDCG@10 on TREC-DL 20, but with only 69\% and 86\% of the re-ranking cost of Fixed-$k$ (1000), respectively.
    \item In the other scenarios where efficiency received more attention ($\beta$=1/2 and $\eta$=0.5/0.6), supervised methods do not show an obvious advantage than unsupervised counterparts. 
    E.g., while AttnCut ($\beta$=2) and MtCut ($\beta$=2) manage to achieve nDCG@10 values comparable to Fixed-$k$ (100) using roughly half the re-ranking cost on TREC-DL 20, Greedy-$k$ ($\beta$=2) attains very similar results as AttnCut and MtCut.
    \item Moreover, as illustrated in Figure~\ref{fig:bm25_rankllama_dl20}, other potential fixed ranking depths (excluding 10, 20, 100, 200 and 1000) can yield results comparable to those of supervised methods across all scenarios.
\end{enumerate*}

Second, in scenarios balancing efficiency and effectiveness or prioritizing efficiency, distribution-based supervised \ac{RLT} methods (Choppy, AttnCut, and MtCut) outperform the sequential labeling-based method (BiCut); however, in the scenario emphasizing effectiveness, BiCut shows a slight advantage.
For instance, as shown in Figure~\ref{fig:bm25_rankllama_dl20}, BiCut incurs lower costs to achieve nDCG@10 comparable to distribution-based methods when effectiveness is emphasized; however,  in other scenarios ($\beta$=1/2 and $\eta$=0.5/0.6), the point denoting BiCut is below the dashed line denoting potential fixed re-ranking depths, while the points representing other supervised methods are on the line, indicating a worse effectiveness/efficiency balance achieved by BiCut.

Third, the supervised method (MtCut) learning \ac{RLT} in a multi-task manner does not show a clear advantage over other supervised methods across all three trade-offs.

\begin{figure*}[!t]
  \centering
  \includegraphics[width=1\linewidth]{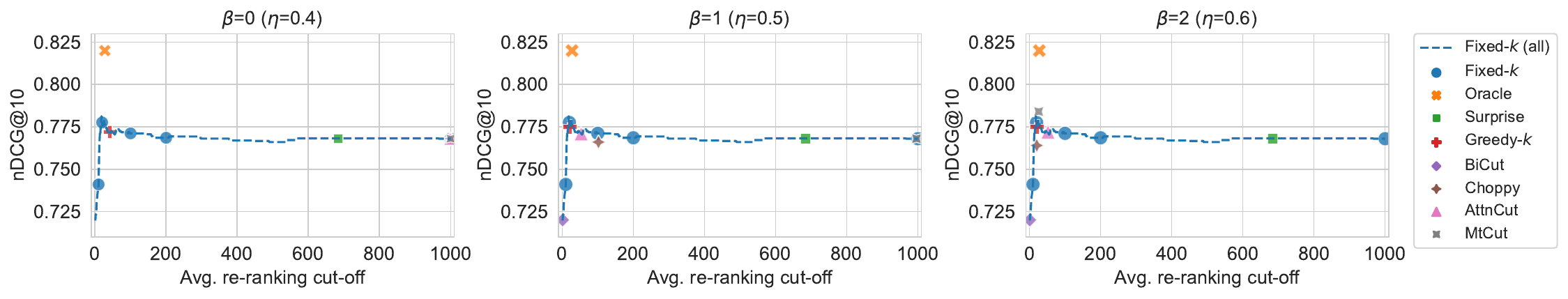}
  \caption{A comparison of RLT methods in predicting re-ranking cut-off points for Splade++--RankLLaMA on TREC-DL 20.}
    \label{fig:splade_rankllama_dl20}
\end{figure*}

\begin{table}[!t]
\centering
\caption{
Comparison of \ac{RLT} methods in predicting re-ranking cut-off points for SPLADE++--RankLLaMA on TREC-DL 19 and 20.
Query latency measured in seconds.
The best nDCG@10 value in each column is marked in \textbf{bold}, and the second best is \underline{underlined}.
Significant differences with Fixed-$k$ (20) are marked with $^*$ (paired t-test, $p < 0.05$).
}
\label{tab:splade_rankllama}
\setlength{\tabcolsep}{.6mm}
\resizebox{\columnwidth}{!}{
\begin{tabular}{l@{~}rcr rcr} 
\toprule
 \multirow{2}{*}{Method} & \multicolumn{3}{c}{TREC-DL 19} & \multicolumn{3}{c}{TREC-DL 20} \\
  \cmidrule(lr){2-4} \cmidrule(lr){5-7}
& \multicolumn{1}{c}{Avg. k} & \multicolumn{1}{c}{nDCG@10}     & \multicolumn{1}{c}{Lat.}   & \multicolumn{1}{c}{Avg. k}   & \multicolumn{1}{c}{nDCG@10}   & \multicolumn{1}{c}{Lat.}      \\
\midrule
w/o re-ranking &  -  &  \phantom{0}0.731$^*$  & -  &  -   & \phantom{0}0.720$^*$    & -  \\
  \midrule
 Fixed-$k$ (10) & 10  & \phantom{0}0.740$^*$   & 0.30  & 10  & \phantom{0}0.741$^*$    & 0.30  \\
 Fixed-$k$ (20) & 20  & \underline{0.773}   & 0.60  & 20  & \underline{0.778}   & 0.60   \\
 Fixed-$k$ (100) & 100  & 0.769   & 2.98   & 100  & 0.771   & 2.98 \\
 Fixed-$k$ (200) & 200   & 0.769   & 5.95  & 200  & 0.769   & 5.96 \\
 Fixed-$k$ (1000) & 1000    & 0.768   & 29.77  & 1000  & 0.768   & 29.79  \\
 Surprise &  702  & 0.766   & 20.90  &  684 & 0.768   & 20.38 \\
 \midrule
Greedy-$k$ ($\beta$=0)  & 65 &  0.770  & 1.94 & 42  &  0.772   & 1.25     \\
BiCut ($\eta$=0.4)  & 1000     & 0.768     & 29.77     &  1000 & 0.768        &  29.79  \\
Choppy ($\beta$=0)  & 904     & 0.768     &  26.92    & 1000  & 0.768        & 29.79  \\
AttnCut ($\beta$=0) & 999     & 0.768      & 29.74     &  999  &  0.768       &  29.76 \\
MtCut ($\beta$=0)  & 999     & 0.768     & 29.74     & 1000  & 0.768        & 29.79   \\
\midrule
%$\alpha$=-0.001 &  &  &  & & & \\
Greedy-$k$ ($\beta$=1)  & 65 & 0.770      & 1.94 & 21 &  0.775    &  0.63    \\
BiCut ($\eta$=0.5)  &  2    & \phantom{0}0.731$^*$     & 0.06     &  1  & \phantom{0}0.720$^*$    & 0.00   \\
Choppy ($\beta$=1) & 68     & 0.771     & 2.03     & 102  &  0.766       &  3.03  \\
AttnCut ($\beta$=1) & 46     & 0.771     & 1.38     & 53  & 0.771        &  1.58  \\
MtCut ($\beta$=1) & 120     & \textbf{0.775}     & 3.56     & 996  & 0.768        &  29.67 \\
%\midrule
%$\alpha$=-0.01 &  &  &  & & & \\
%Greedy-$k$ ($\beta$=1)  & 18 & 0.769      &  & 21 & 0.775    &      \\
%Greedy-$k$ ($\beta$=2)   & 18 &  0.769      &  & 21 & 0.775    &      \\
\midrule
%BiCut ($\alpha$=0.45)  & 1000.0     & 0.768     &      & 1000.0  & 0.768    &   \\
%BiCut ($\alpha$=0.55)  & 2.0     & 0.731     &      & 1.0  &  0.720    &   \\

%BiCut ($\alpha$=0.65)  & 0.0     &  0.731    &      &  1.0 &  0.720    &   \\
%BiCut ($\alpha$=0.70)  & 0.0     &  0.731    &      &  1.0 &  0.720    &   \\
%BiCut ($\alpha$=0.75)  &  0.0    & 0.731     &      &  1.0 & 0.720     &  \\
%BiCut ($\alpha$=0.80) & 0.0     & 0.731     &      &  1.0 & 0.720     &   \\
Greedy-$k$ ($\beta$=2)   & 18 & 0.769      & 0.54 &  21 &  0.775   &  0.63    \\
BiCut ($\eta$=0.6)  & 2    & \phantom{0}0.731$^*$     & 0.06     & 1  & \phantom{0}0.720$^*$     & 0.00  \\
%$\alpha$=-0.001 &  &  &  & & & \\
Choppy ($\beta$=2) & 1     &  \phantom{0}0.731$^*$    &  0.00     & 21  &  0.764       &  0.61  \\
AttnCut ($\beta$=2)  &  24    &  \phantom{0}0.758$^*$    &  0.70    & 52  &  0.772       & 1.55   \\
MtCut ($\beta$=2) & 20     &  \phantom{0}0.756$^*$    & 0.59     & 26  &  \textbf{0.784}      & 0.77   \\
%\midrule
%$\alpha$=-0.01 &  &  &  & & & \\
%Choppy ($\beta$=1) &  49.56    & 0.767     &      & 31.63  & 0.770        &   \\
%AttnCut ($\beta$=1) & 17.02     & 0.745     &      & 7.0  & 0.736        &   \\
%MtCut ($\beta$=1)  & 20.98     & 0.761     &      & 969.87  & 0.768        &   \\
%Choppy ($\beta$=2)  & 21.07     & 0.762     &      & 26.96  & 0.772        &   \\
%AttnCut ($\beta$=2)  & 9.79     & 0.746     &      & 4.0 &  0.728       &   \\
%MtCut ($\beta$=2)  & 19.67     & 0.747     &      & 22.02  & 0.771        &   \\
%\midrule
%$\alpha$=0.05 &  &  &  & & & \\
%Choppy ($\beta$=1)  &  28.91    & 0.759     &      &  10.52 &  0.752       &   \\
%AttnCut ($\beta$=1) &  7.35    &  0.733    &      &  4.0 & 0.728        &   \\
%MtCut ($\beta$=1)  & 62.44     & 0.745     &      & 1000  & 0.768        &   \\
%Choppy ($\beta$=2) &  1.58    & 0.734     &      &  1.0  &  0.720       &   \\
%AttnCut ($\beta$=2)  & 4.88     & 0.733     &      & 3.0  &  0.723       &   \\
%MtCut ($\beta$=2)  & 2.12     & 0.734     &      & 38.0  &  0.716       &   \\
\midrule
 Oracle & 17   & \phantom{0}0.810$^*$   & 0.52   & 28  & \phantom{0}0.820$^*$    & 0.82 \\
\bottomrule
\end{tabular}
}
\end{table}

\if0
\begin{figure*}[!t]
  \centering
  \includegraphics[width=.9\linewidth]{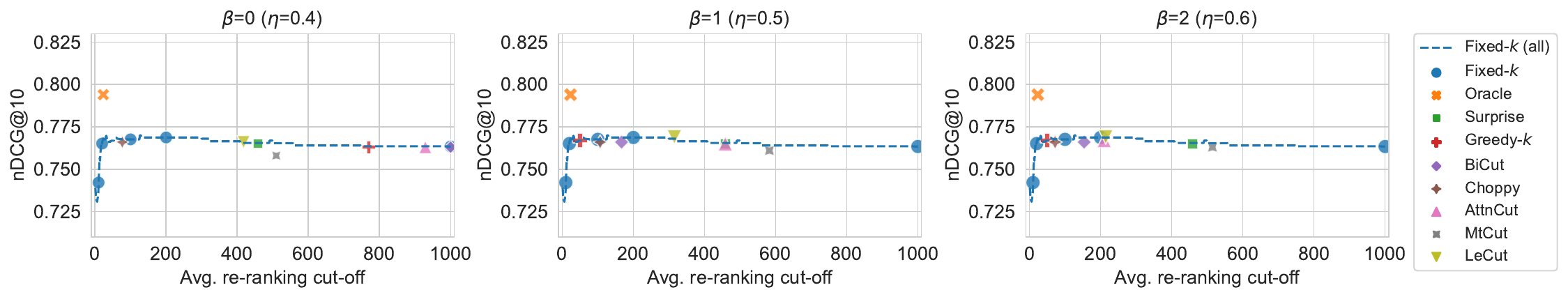}
  \caption{RepLLaMA--RankLLaMA on TREC-DL 19.}
    \label{fig:repllama_rankllama_dl19}
\end{figure*}
\fi

\begin{figure*}[!t]
  \centering
  \includegraphics[width=1\linewidth]{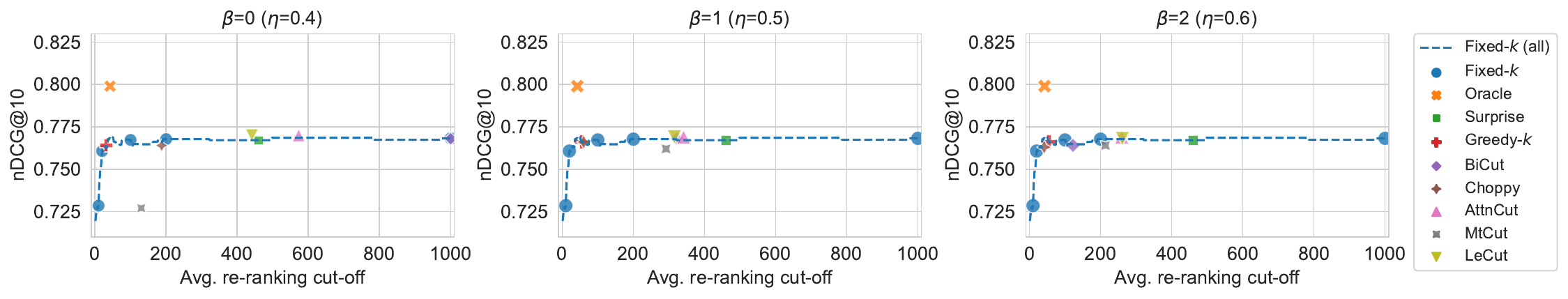}
  \caption{A comparison of RLT methods in predicting re-ranking cut-off points for RepLLaMA--RankLLaMA on TREC-DL 20.}
    \label{fig:repllama_rankllama_dl20}
\end{figure*}

\begin{table}[!t]
\centering
\caption{
Comparison of \ac{RLT} methods in predicting cut-off points for RepLLaMA--RankLLaMA on TREC-DL 19 and 20.
Query latency measured in seconds.
The best nDCG@10 value in each column is marked in \textbf{bold}, and the second best is \underline{underlined}.
Significant differences with Fixed-$k$ (20) are marked with $^*$ (paired t-test, $p < 0.05$).
}
\label{tab:repllama_rankllama}
\setlength{\tabcolsep}{1mm}
\resizebox{\columnwidth}{!}{
\begin{tabular}{l rcr rcr} 
\toprule
 \multirow{2}{*}{Method} & \multicolumn{3}{c}{TREC-DL 19} & \multicolumn{3}{c}{TREC-DL 20} \\
  \cmidrule(lr){2-4} \cmidrule(lr){5-7}
& \multicolumn{1}{c}{Avg. k} & \multicolumn{1}{c}{nDCG@10}     & \multicolumn{1}{c}{Lat.}   & \multicolumn{1}{c}{Avg. k}   & \multicolumn{1}{c}{nDCG@10}   & \multicolumn{1}{c}{Lat.}      \\
\midrule
w/o re-ranking & - & \phantom{0}0.738$^*$      & -  & - & \phantom{0}0.720$^*$    & -   \\ 
  \midrule
  Fixed-$k$ (10) & 10  & \phantom{0}0.742$^*$   & 0.30   & 10   &  \phantom{0}0.729$^*$    & 0.30   \\
  Fixed-$k$ (20) & 20  & 0.765   & 0.60     & 20    & 0.761     & 0.60    \\
  Fixed-$k$ (100) & 100  & \textbf{0.769}    & 2.98   & 100  &  0.767   & 2.99  \\
  Fixed-$k$ (200) & 200  & \underline{0.768}    &  5.96   & 200   & 0.768  & 5.97  \\
  Fixed-$k$ (1000) & 1000  & 0.763    & 29.81    & 1000   & 0.768  & 29.86  \\
 Surprise & 458     &  0.765     &  13.66 & 460   &  0.767     & 13.73  \\
\midrule
Greedy-$k$ ($\beta$=0)  & 770 & 0.763      & 22.95  & 31 & 0.764    &  0.93    \\
BiCut ($\eta$=0.4)  & 1000   & 0.763    & 29.81     &  1000  & 0.768        & 29.86   \\
Choppy ($\beta$=0)  & 77     &  0.766    & 2.29     & 187  &  0.764       & 5.60   \\
AttnCut ($\beta$=0) &  929    & 0.763     & 27.69     & 573   & \textbf{0.770}        &  17.10 \\
MtCut ($\beta$=0)  & 510     &  0.758    & 15.19     & 130  &  \phantom{0}0.727$^*$       & 3.88   \\
LeCut ($\beta$=0)  &  418    & 0.766     & 12.45     & 441  & \textbf{0.770}        & 13.17   \\
\midrule
%$\alpha$=-0.001 &  &  &  & & & \\
Greedy-$k$ ($\beta$=1)  & 50 & 0.767      & 1.49  & 55  &  0.766   &  1.64    \\
BiCut ($\eta$=0.5)  & 167    &  0.766   & 4.97     &  323 & 0.768     & 9.63  \\
%\midrule
%$\alpha$=-0.01 &  &  &  & & & \\
%Greedy-$k$ ($\beta$=1)  & 40 & 0.766      &  & 31  &  0.764   &      \\
%Greedy-$k$ ($\beta$=2)   & 20 & 0.765      &  & 31 &  0.764   &      \\
Choppy ($\beta$=1) &  107    &  0.766    &  3.18    &  61 &  0.766       &  1.81  \\
AttnCut ($\beta$=1) & 458     & 0.765     &  13.67    & 341  & \underline{0.769}       & 10.19  \\
MtCut ($\beta$=1) & 583     &  0.761    & 17.37     & 292  &  0.762       &  8.71  \\
LeCut ($\beta$=1)  &  315    & \textbf{0.769}     & 9.40     & 316  &   \underline{0.769}      & 9.42   \\
\midrule
%BiCut ($\alpha$=0.45)  & 198.91    & 0.766    &      & 179.96  & 0.767     &   \\
%BiCut ($\alpha$=0.55)  & 133.09     & 0.766     &      & 192.28  & 0.768     &   \\
Greedy-$k$ ($\beta$=2)   & 50 & 0.767      & 1.49 & 55 &  0.766   & 1.64     \\
BiCut ($\eta$=0.6)  & 154     & 0.766     & 4.58     & 122  & 0.764     & 3.65   \\
%BiCut ($\alpha$=0.65)  &  90.42    & 0.767     &      & 0.0  & 0.720     &   \\
%BiCut ($\alpha$=0.70)  & 29.63     & 0.766     &      & 0.0  &  0.720    &   \\
%BiCut ($\alpha$=0.75)  &  1    &  0.738    &      &   &      &  \\
%BiCut ($\alpha$=0.80) &  1    &  0.738    &      &   &      &   \\
Choppy ($\beta$=2) &  72    &  0.766    &  2.15    &  41 & 0.763        &  1.24  \\
AttnCut ($\beta$=2) & 210     & 0.767     & 6.26     &  259 &  \underline{0.769}       & 7.74   \\
MtCut ($\beta$=2) & 515     & 0.763     & 15.34     & 214  & 0.764        & 6.38   \\
LeCut ($\beta$=2)  & 214     & \textbf{0.769}     & 6.39     &  261  &  0.768      & 7.80   \\
%$\alpha$=-0.001 &  &  &  & & & \\
\midrule
%$\alpha$=-0.01 &  &  &  & & & \\
%Choppy ($\beta$=1) & 40.91     & 0.766    &      &  33.69 & 0.764        &   \\
%AttnCut ($\beta$=1) & 13.67   &  0.750    &      & 83.41  & 0.748        &   \\
%MtCut ($\beta$=1)  &  53.30    &  0.741    &      &  42.04 &  0.725       &   \\
%LeCut ($\beta$=1)  & 55.67     & 0.767     &      & 75.07  &  0.767       &   \\
%Choppy ($\beta$=2) & 26.74     & 0.769     &      & 33.00  &  0.764       &   \\
%AttnCut ($\beta$=2) & 29.14     &  0.763    &      & 22.70  & 0.753        &   \\
%MtCut ($\beta$=2)  & 27.91     &  0.741    &      & 40.78  & 0.730        &   \\
%LeCut ($\beta$=2)  & 36.70     & 0.769     &      & 42.72  &  0.765       &   \\
%\midrule
%$\alpha$=0.05 &  &  &  & & & \\
%Choppy ($\beta$=1) &  22.0    &  0.767    &      & 22.00  & 0.761        &   \\
%AttnCut ($\beta$=1) & 7.74     & 0.733     &      &  13.09  &  0.729       &   \\
%MtCut ($\beta$=1)  &  56.58    &  0.737    &      &   1.06 &  0.719       &   \\
%LeCut ($\beta$=1)  & 78.33     &  0.766    &      & 58.89  & 0.767        &   \\
%Choppy ($\beta$=2) &  16.09    & 0.754     &      & 15.0 &   0.749       &   \\
%AttnCut ($\beta$=2)  &  7.16    & 0.733     &      &  14.39 & 0.730        &   \\
%MtCut ($\beta$=2)  &  13.23    &  0.738    &      & 1.41  &  0.719       &   \\
%LeCut ($\beta$=2)  & 34.12     &   0.760   &      & 32.56  & 0.764        &   \\
%\midrule
  Oracle & 23   & \phantom{0}0.794$^*$     & 0.70  & 42  &  \phantom{0}0.799$^*$   & 1.27 \\
\bottomrule
\end{tabular}
}
\end{table}

\subsection{The impact of retriever types on \ac{RLT}}
\label{res:rq2}

To answer \ref{RQ2}, we assess \ac{RLT} methods (optimized for the three effectiveness/efficiency trade-offs) in the context of an \ac{LLM}-based re-ranker (RankLLaMA~\citep{ma2023fine}) with other novel retrievers; we explore learned sparse (SPLADE++~\citep{formal2022distillation}) and dense~(RepLLaMA~\citep{ma2023fine}) retrievers. %\looseness=-1
We present the results for SPLADE++--RankLLaMA and RepLLaMA--RankLLaMA on TREC-DL 19 and 20 in Table~\ref{tab:splade_rankllama} and \ref{tab:repllama_rankllama}, respectively.
we plot both pipelines' results for TREC-DL 20 in Figure~\ref{fig:splade_rankllama_dl20} and \ref{fig:repllama_rankllama_dl20}.
Note that LeCut is only compatible with pipelines featuring a dense retriever.
%Figure~\ref{fig:splade_rankllama_dl19} and \ref{fig:splade_rankllama_dl20}, as well as Table~\ref{tab:splade_rankllama}; we show the results for RepLLaMA--RankLLaMA on TREC-DL 19 and 20 in Figure~\ref{fig:repllama_rankllama_dl19} and \ref{fig:repllama_rankllama_dl20}, as well as Table~\ref{tab:repllama_rankllama}.
We have four observations.

First, different from the findings in Section~\ref{res:rq1}, the unsupervised method Fixed-$k$ (20) consistently achieves the best effectiveness/efficiency trade-off compared to supervised methods for both pipelines across all scenarios.
Although some supervised methods achieve higher nDCG@10 values than Fixed-$k$ (20), the nDCG@10 improvement is not statistically significant, e.g., for SPLADE++--RankLLaMA, MtCut ($\beta$=2) outperform Fixed-$k$ (20) by 0.006 in terms of nDCG@10 at a comparable cost on TREC-DL 20; however, this difference is too marginal.
We believe Fixed-$k$ (20) performs well due to the superior retrieval capabilities of SPLADE++ and RepLLaMA.
Both tend to retrieve more relevant items at the top ranks, making a shallow fixed re-ranking cutoff both effective and efficient.

%unsupervised methods exhibit a marked advantage over supervised methods for both pipelines.
%E.g., Fixed-$k$ (10) achieves the best nDCG@10 values while maintaining lowest re-ranking cost for SPLADE++--RankLLaMA on TREC-DL 19 and TREC-DL 20 (except for $\beta$=2); Fixed-$k$ (10) and Greedy-$k$ show the same case for RepLLaMA--RankLLaMA on TREC-DL 19 and 20.

Second, similar to Section~\ref{res:rq1}, distribution-based supervised methods typically offer a better effectiveness/efficiency trade-off in re-ranking than the sequential labeling-based method (BiCut). 
E.g., for SPLADE++--RankLLaMA, BiCut predicts depths that are either too shallow (1 or 2) or too deep (1000); for RepLLaMA--RankLLaMA, certain distribution-based methods (e.g., Choppy) achieve similar nDCG@10 values to BiCut but at a reduced re-ranking cost.

Third, different from Section~\ref{res:rq1}, the supervised method (MtCut), which learns \ac{RLT} in a multi-task manner, exhibits a superior effectiveness/efficiency trade-off compared to other methods in a specific case; however, this advantage is not consistently observed.
Specifically, as shown in Figure~\ref{fig:splade_rankllama_dl20}, in the scenario emphasizing efficiency, MtCut ($\beta$=2) achieves the highest nDCG@10 value (except for Oracle) for SPLADE++--RankLLaMA on TREC-DL 20 at a notably low cost.
Nevertheless, as depicted in Figure \ref{fig:repllama_rankllama_dl20}, for RepLLaMA--RankLLaMA, the points representing MtCut consistently fall below the dashed line of Fixed-$k$, indicating a worse effectiveness/efficiency trade-off compared to other supervised methods.

%, while maintaining the lowest re-ranking cost for SPLADE++--RankLLaMA on TREC-DL 20 ($\beta=2$); however, MtCut does not keep this trend for RepLLaMA--RankLLaMA.
Fourth, LeCut utilizes query-item embeddings from RepLLaMA to predict re-ranking cut-off points for RepLLaMA--RankLLaMA, leading to marginal improvements over other supervised methods in nDCG@10. 
%
%As shown in Table~\ref{tab:repllama_rankllama} and Figure~\ref{fig:repllama_rankllama_dl20}, 
These improvements are often too minimal to be significant; moreover, LeCut often attains these marginal improvements at the cost of efficiency.
E.g., LeCut ($\beta$=0) achieves the highest nDCG@10 value of 0.770 on TREC-DL 20, outperforming that of Choppy ($\beta$=0) by 0.006, yet at more than double the cost of Choppy.

%better nDCG@10 values than other methods
%E.g., LeCut ($\beta=0$) on TREC-DL 20 achieves the highest nDCG@10 value (0.770).
%shows better nDCG@10 values when optimized for prioritizing effectiveness.

\if0
\begin{figure*}[!t]
  \centering
  \includegraphics[width=.9\linewidth]{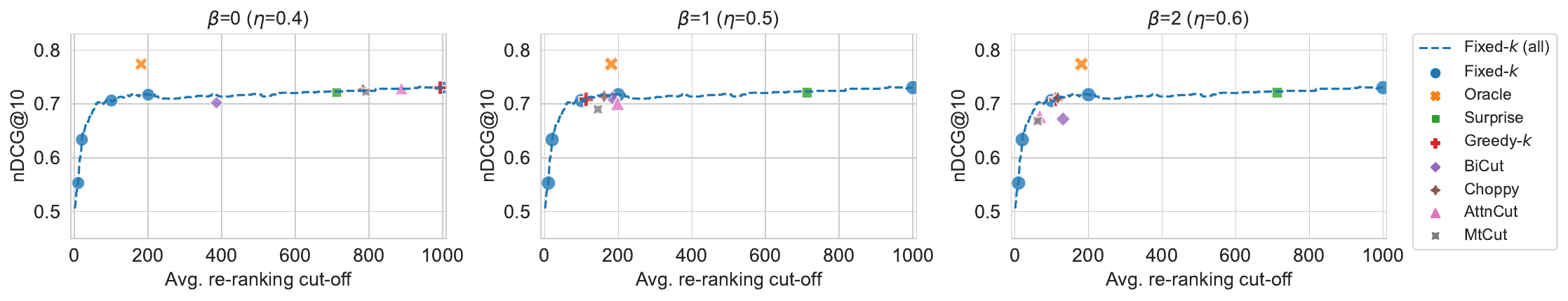}
  \caption{BM25--monoT5 on TREC-DL 19.}
    \label{fig:bm25_monot5_dl19}
\end{figure*}
\fi

\begin{figure*}[ht]
  \centering
  \includegraphics[width=1\linewidth]{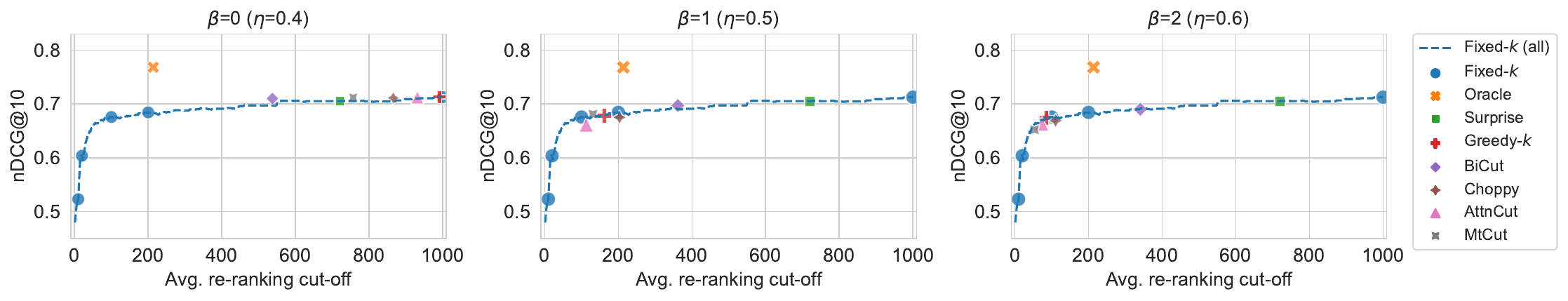}
  \caption{A comparison of RLT methods in predicting re-ranking cut-off points for BM25--monoT5 on TREC-DL 20.}
    \label{fig:bm25_monot5_dl20}
\end{figure*}

\subsection{\ac{RLT} for pre-trained LM-based re-ranking}

To answer \ref{RQ3}, we evaluate \ac{RLT} methods in the context of a pre-trained language model-based re-ranker (monoT5~\citep{nogueira2020document}) with a lexical retriever (BM25).
Note that using monoT5~\citep{nogueira2020document} to re-rank the retrieved list returned by RepLLaMA~\citep{ma2023fine} and Splade++~\citep{formal2022distillation} yields worse results; hence we only consider the pipeline of BM25--monoT5.
We report the raw result numbers on TREC-DL 19 and 20 in Table~\ref{tab:bm25_monot5}.
We also plot the results on TREC-DL 20 in Figure~\ref{fig:bm25_monot5_dl20}.

Generally, the findings obtained with pre-trained language model-based and an \ac{LLM}-based re-rankers (see Section~\ref{res:rq1}) are similar.
Specifically, 
\begin{enumerate*}[label=(\roman*)]
    \item compared to unsupervised ones, supervised methods only show an advantage in the scenario emphasizing effectiveness, where supervised methods achieve better re-ranking quality at a lower cost; e.g., BiCut ($\eta$=0.4) achieves an nDCG@10 value comparable to that of Fixed-$k$ (1000) while incurring only half the cost on TREC-DL 20;  however, similar to Section~\ref{res:rq1}, potential fixed re-ranking depths (excluding 10, 20, 100, 200, and 1000) can still yield results that are very similar to those obtained with supervised methods (see Figure~\ref{fig:bm25_monot5_dl20});
    \item distribution-based supervised methods perform better in scenarios balancing efficiency and effectiveness or prioritizing efficiency, while the sequential labeling-based method (BiCut) shows a slight advantage in the scenario emphasizing effectiveness; and
    \item the supervised method (MtCut), which learns \ac{RLT} in a multi-task manner, does not consistently show a clear advantage over other supervised methods across all three trade-offs.
\end{enumerate*}

%first, supervised \ac{RLT} methods only show an advantage in terms of re-ranking e;
%
%second, the most distribution-based supervised \ac{RLT} methods outperform the sequential labeling-based method (BiCut) in terms of effectiveness and efficiency; and
%
%third, optimizing \ac{RLT} in a multi-task manner does not lead to consistent improvement in effectiveness and efficiency.

%We observe up to 18× speedup without degradation in terms of MRR@10

\begin{table}[t]
\centering
\caption{
A comparison of \ac{RLT} methods in predicting re-ranking cut-off points for BM25--monoT5 on TREC-DL 19 and 20.
Query latency measured in seconds.
The best nDCG@10 value in each column is marked in \textbf{bold}, and the second best is \underline{underlined}.
Significant differences with Fixed-$k$ (100), Fixed-$k$ (200) and Fixed-$k$ (1000) are marked with $^*$, $^\S$ and $^\dagger$, respectively (paired t-test, $p < 0.05$).
}   
\label{tab:bm25_monot5}
\setlength{\tabcolsep}{1mm}
\resizebox{\columnwidth}{!}{
\begin{tabular}{l rcr rcr} 
\toprule
 \multirow{2}{*}{Method} & \multicolumn{3}{c}{TREC-DL 19} & \multicolumn{3}{c}{TREC-DL 20}  \\
  \cmidrule(lr){2-4} \cmidrule(lr){5-7}
& \multicolumn{1}{c}{Avg. k} & \multicolumn{1}{c}{nDCG@10}     & \multicolumn{1}{c}{Lat.}  & \multicolumn{1}{c}{Avg. k}   & \multicolumn{1}{c}{nDCG@10}    & \multicolumn{1}{c}{Lat.}     \\
\midrule
w/o re-ranking & -  &  \phantom{000}0.506$^*$$^\S$$^\dagger$    & -   & -   & \phantom{000}0.480$^*$$^\S$$^\dagger$  &  -  \\
  \midrule
  Fixed-$k$ (10)  & 10    & \phantom{000}0.553$^*$$^\S$$^\dagger$     & 0.14     & 10  & \phantom{000}0.523$^*$$^\S$$^\dagger$        &  0.14 \\
  Fixed-$k$ (20) & 20   & \phantom{000}0.634$^*$$^\S$$^\dagger$      & 0.27    & 20  &  \phantom{000}0.604$^*$$^\S$$^\dagger$       & 0.27 \\
  Fixed-$k$ (100) & 100    & 0.706     & 1.37    & 100  & \phantom{0}0.676$^\dagger$        & 1.37  \\
  Fixed-$k$ (200) & 200   & 0.717      &  2.73   & 200  & \phantom{0}0.684$^\dagger$        & 2.73 \\
 Fixed-$k$ (1000)  & 1000      & \textbf{0.730}     &  13.66    & 1000   & \phantom{00}\textbf{0.713}$^*$$^\S$        & 13.66 \\
%Greedy-$k$  &      &       &  &    &       &      \\
Surprise &  712    & 0.721       & 9.73 & 721   & \phantom{00}0.705$^*$$^\S$      & 9.84 \\
 \midrule
Greedy-$k$ ($\beta$=0)  & 993 & \textbf{0.730}   & 13.57 & 991 &  \phantom{00}\textbf{0.713}$^*$$^\S$   & 13.54   \\
BiCut ($\eta$=0.4)  & 386     &  0.702    & 5.27     & 538  & \phantom{00}0.710$^*$$^\S$        & 7.34 \\
Choppy ($\beta$=0)  & 784     & 0.727     &  10.70    & 866  &  \phantom{00}0.711$^*$$^\S$       & 11.82  \\
AttnCut ($\beta$=0) & 888     & \underline{0.729}      &  12.13   & 931  & \phantom{00}\underline{0.712}$^*$$^\S$        & 12.72   \\
MtCut ($\beta$=0)  & 791     & 0.722     & 10.81    & 757  &  \phantom{00}\underline{0.712}$^*$$^\S$       & 10.34  \\
\midrule
%$\alpha$=-0.001 &  &  &  & & & \\
Greedy-$k$ ($\beta$=1)  & 112 & 0.709      & 1.53  & 162  & \phantom{0}0.678$^\dagger$    & 2.21   \\
BiCut ($\eta$=0.5)  & 184     & 0.711     & 2.51    & 362  & 0.697     & 4.94   \\
Choppy ($\beta$=1) & 161     &  0.714    & 2.20     & 203  & \phantom{0}0.675$^\dagger$        & 2.78  \\
AttnCut ($\beta$=1) & 198     & 0.701     & 2.70    & 113  &  \phantom{0}0.661$^\dagger$       & 1.54  \\
MtCut ($\beta$=1) &  145    & \phantom{00}0.690$^\S$$^\dagger$     & 1.97    &  131 & 0.681        &  1.79  \\
\midrule
Greedy-$k$ ($\beta$=2)   & 112 & 0.709       & 1.53 & 86 & \phantom{0}0.674$^\dagger$    & 1.17     \\
BiCut ($\eta$=0.6)  & 131     & \phantom{000}0.672$^*$$^\S$$^\dagger$     & 1.79     & 341  & 0.690     & 4.66   \\
Choppy ($\beta$=2)  & 117     & 0.711     & 1.59     & 111  & \phantom{0}0.669$^\dagger$        & 1.51   \\
AttnCut ($\beta$=2)  & 68     & \phantom{000}0.677$^*$$^\S$$^\dagger$     &  0.92    & 73  & \phantom{0}0.663$^\dagger$        & 1.00   \\
MtCut ($\beta$=2) & 62     & \phantom{000}0.668$^*$$^\S$$^\dagger$    &  0.85    & 54  & \phantom{0}0.652$^\dagger$        & 0.73   \\
%$\alpha$=-0.01 &  &  &  & & & \\
%Greedy-$k$ ($\beta$=1)  & 53 & 0.690      &  & 45 & 0.654     &      \\
%Greedy-$k$ ($\beta$=2)   & 29 &  0.658     &  & 30 & 0.627    &      \\
%\midrule

%BiCut ($\alpha$=0.45)  & 168.40     & 0.688     &      & 501.56  & 0.713     &   \\

%BiCut ($\alpha$=0.55)  & 130.49     & 0.669     &      & 412.59  & 0.705     &   \\

%BiCut ($\alpha$=0.65)  &  99.09    & 0.665     &      & 204.69  & 0.675     &   \\
%BiCut ($\alpha$=0.70)  & 34.67     &  0.579    &      & 124.74  & 0.645     &   \\
%BiCut ($\alpha$=0.75)  & 10.28     & 0.533     &      & 41.74  &  0.524    &  \\
%BiCut ($\alpha$=0.80) & 0.0     & 0.506     &      & 0.0   & 0.480     &   \\
%$\alpha$=-0.001 &  &  &  & & & \\
\midrule
%$\alpha$=-0.01 &  &  &  & & & \\
%Choppy ($\beta$=1) &  39.65    &  0.684    &      & 51.81 & 0.660        &   \\
%AttnCut ($\beta$=1) & 48.19     & 0.633     &      & 39.54  & 0.624        &   \\
%MtCut ($\beta$=1)  & 41.09     & 0.599     &      & 27.26  &  0.603       &   \\
%Choppy ($\beta$=2)  & 30.84     & 0.660     &      & 25.41  &  0.594       &   \\
%AttnCut ($\beta$=2) & 40.84     & 0.604     &      & 26.65 &  0.611       &   \\
%MtCut ($\beta$=2)  & 37.02     & 0.573     &      &  20.00 & 0.586        &   \\
%\midrule
%$\alpha$=0.05 &  &  &  & & & \\
%Choppy ($\beta$=1)  & 21.44     & 0.638     &      & 20.63  & 0.594        &   \\
%AttnCut ($\beta$=1) & 9.72     & 0.547     &      &  9.65 & 0.531         &   \\
%MtCut ($\beta$=1)  &  32.72    & 0.559     &      & 9.26  & 0.551        &   \\
%Choppy ($\beta$=2)  & 14.00     &  0.599    &      & 17.48  &  0.591       &   \\
%AttnCut ($\beta$=2)  & 8.28     & 0.541     &      & 8.41  & 0.518        &   \\
%MtCut ($\beta$=2)  & 8.44     & 0.548     &      & 7.48  &  0.534       &   \\
%\midrule
  Oracle &  181    &  \phantom{000}0.774$^*$$^\S$$^\dagger$    & 2.47     & 214  & \phantom{000}0.768$^*$$^\S$$^\dagger$        &  2.92 \\
\bottomrule
\end{tabular}
}
\end{table}

\begin{figure*}[!t]
    \centering
    \begin{subfigure}{0.5\columnwidth}
        \includegraphics[width=\linewidth]{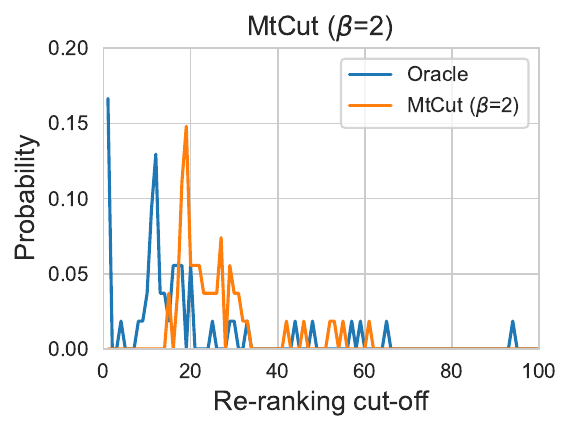}
        \vspace*{-7mm}
        \caption{SPLADE++–RankLLaMA}
        \label{fig:mmoecut-splade-rankllama}
    \end{subfigure}
    \begin{subfigure}{0.5\columnwidth}
        \includegraphics[width=\linewidth]{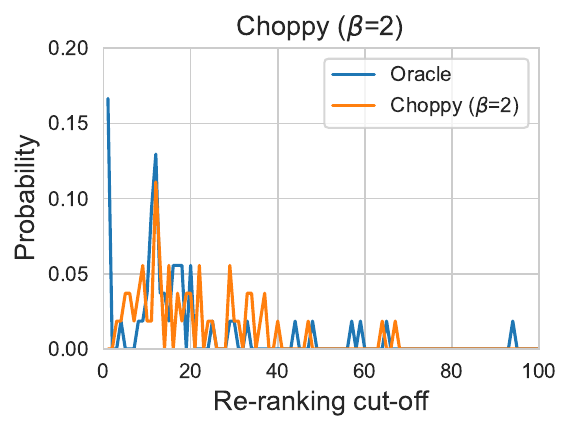}
        \vspace*{-7mm}
        \caption{SPLADE++–RankLLaMA}
        \label{fig:choppy-splade-rankllama}
    \end{subfigure}
    \begin{subfigure}{0.5\columnwidth}
        \includegraphics[width=\linewidth]{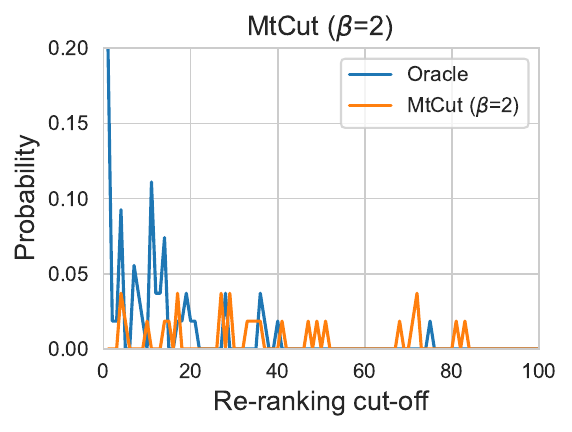}
        \vspace*{-7mm}
        \caption{RepLLaMA--RankLLaMA}
        \label{fig:mmoecut-repllama-rankllama}
    \end{subfigure}
    \begin{subfigure}{0.5\columnwidth}
        \includegraphics[width=\linewidth]{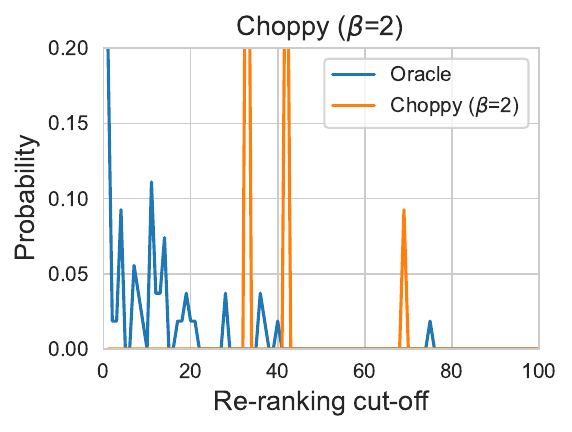}
        \vspace*{-7mm}
        \caption{RepLLaMA--RankLLaMA}
        \label{fig:choppy-repllama-rankllama}
    \end{subfigure}
    \caption{
    The distribution of re-ranking cut-off points on TREC-DL 20. 
    }
    \label{fig:error}
    \vspace*{1mm}
\end{figure*}

\subsection{Error analysis}
\label{sec:error}
To understand the reasons behind the less-than-ideal performance of supervised \ac{RLT} methods, we compare distributions of re-ranking cut-off points predicted by Oracle with those predicted by a supervised method.
We consider two relatively effective methods,\footnote{Due to space limitations, we provide error analysis for all methods in our repository.} MtCut ($\beta$=2) and Choopy ($\beta$=2), for pipelines featuring novel retrievers, SPLADE++ and RepLLaMA on TREC-DL 20; see Figure~\ref{fig:error}.
%
%We have two observations.
First, both methods fail to predict a re-ranking cut-off of zero.
For both pipelines, around 20\% of queries do not need re-ranking.
Thus, enhancing supervised \ac{RLT} methods' ability to predict when re-ranking is unnecessary is crucial.
Second, both methods perform worse when truncating RepLLaMA's retrieved lists (see Figure~\ref{fig:mmoecut-repllama-rankllama} and \ref{fig:choppy-repllama-rankllama}) compared to SPLADE++ (see Figure \ref{fig:mmoecut-splade-rankllama} and \ref{fig:choppy-splade-rankllama}).
Especially, Choppy ($\beta$=2) seems to underfit for RepLLaMA~(see \ref{fig:choppy-repllama-rankllama}), suggesting a potential need for more training data.
%
%We leave a more in-depth analysis to future work.

%% file: Sections/06_Related_Work.tex
\section{Related Work}
%Our work is relevant to two strands of research: \acf{RLT} and \acp{LLM} as re-rankers.

\header{Ranked list truncation}
\Acf{RLT} is also known as query cut-off prediction~\citep{cohen2021not,lesota2021modern}.
For a query and a ranked list of documents, the \ac{RLT} task is to predict the number of items in the ranked list that should be returned, to optimize a user-defined metric~\citep{bahri2023surprise}.
%
%In other words, this task aims to strike a balance between the overall utility of search results and the user cost of processing search results~\citep{wang2022mtcut}.
%
The task can potentially benefit \ac{IR} applications where it is money- and time-consuming to review a returned item, e.g., in patent search~\citep{lupu2013patent} and legal search~\citep{wang2022mtcut,tomlinson2007overview}.  
Early work is mainly assumption-based (hence non-neural-based).
This kind of research focuses on modeling score distributions by fitting prior distributions to them~\citep{arampatzis2009stop,manmatha2001modeling}, which helps identify the best cut-off.
However, prior assumptions on score distributions do not always hold as retrieval settings change~\citep{wang2022mtcut,lien2019assumption}; hence we do not study this line of studies in our work.
Assumption-free methods, on the other hand, learn to predict the truncation position during training and do not rely on a prior assumption.
We have already introduced those methods (BiCut, Choppy, AttnCut, MtCut and LeCut) in Section~\ref{sec:setup}.

%\citet{lien2019assumption} use a bidirectional \ac{LSTM} to make a binary prediction (continue or truncate) at each position over a ranked list with a loss function that serves as a proxy to an IR metric as a target metric (e.g., F1).
%
%\citet{bahri2020choppy} use a transformer encoder~\citep{vaswani2017attention} to predict a probability distribution over all candidate cut-off positions, training the model by a loss function that optimizes the expected value of a target metric.
%
%\citet{wu2021learning} improve the optimization by employing \ac{RAML} to optimize a target metric directly.
%
%\citet{wang2022mtcut} identify the negative effect of retrieval bias on \ac{RLT} and improve the optimization through multi-task learning.
%
%\citet{ma2022incorporating} introduce semantic features and context information from the retrieval model and jointly optimize \ac{RLT} and an external \ac{LtR} model.
%
%Recently, \citet{bahri2023surprise} propose a statistical scoring method based on extreme value theory~\citep{pickands1975statistical} to calibrate scores in a ranked list, leading to improved \ac{RLT} quality. 

\citet{zamani2022stochastic} apply the \ac{RLT} method from \citep{bahri2020choppy} to truncate retrieval lists returned by BM25 for BERT-based re-ranking~\citep{nogueira2019passage}, finding that truncating the retrieval result list to avoid including a large number of non-relevant items in the lower ranks, achieves better re-ranking performance than using fixed cut-offs for all queries.

We differ from \citet{zamani2022stochastic} as we provide a systematic and comprehensive study into the use of \ac{RLT} methods in the context of re-ranking, especially newly emerged \acp{LLM}-based re-ranking.

\if0
\cm{to do: the discussions for the following papers need to be added}
\citep{jiang2022relevance}
\citep{wild2022cost}
\citep{cohen2021not}
\citep{lesota2021modern}
\citep{cohen2022inconsistent}
\fi

%This is due to the number of non-relevant documents that are included by this fixed cutoff value.
%higher retrieval recall (1000) leads to worse performance
% one may want to estimate how far the horizon of the search should be

%\vspace*{-2mm}
\header{Improving neural re-ranking efficiency}
Improving neural re-ranking efficiency has been extensively studied.
There are two ideas to improve the efficiency~\citep{gienapp2022sparse}: 
\begin{enumerate*}[label=(\roman*)]
    \item speed up inference of a neural re-ranker, and
    \item reducing the number of inferences of a neural re-ranker.
\end{enumerate*} 
Approaches to (i) include a simpler re-ranker model~\citep{hofstatter2020interpretable}, distilling knowledge in BERT~\citep{devlin2019bert} into a smaller re-ranker~\citep{gao2020understanding}, pre-computing item representations at indexing time~\citep{macavaney2020efficient}, and early-exiting~\citep{xin2020early,soldaini2020cascade}.
%Early-exiting is to use only a partial model for ``easy'' item~\citep{bruch2023efficient}, which also has been studied in \ac{LtR}~\citep{busolin2021learning,lucchese2020query,cambazoglu2010early}.
%
Studies into (ii) are more related to our work.
It includes \textit{multi-stage re-ranking}~\citep{zhang2021learning,matsubara2020reranking,nogueira2019multi,wang2011cascade} and \textit{candidate pruning}~\citep{li2022certified,nogueira2019multi,culpepper2016dynamic}.
%and \textit{early stopping}~\citep{ying2020adaptive}.

\textit{Multi-stage re-ranking} first exploits faster and less effective re-rankers to discard likely non-relevant items and sends fewer candidate items to more expensive re-rankers in later stages. 
E.g., \citet{zhang2021learning} first use a feature-based \ac{LtR} model to reorder the items returned by BM25 and then send the top-$k$ (applied to all queries) items returned by the faster re-ranker to a BERT re-ranker.

\textit{Candidate pruning} trims the candidate list in the first (or earlier) stage and then forwards the pruned ranked list to the next stage re-ranking.
\citet{wang2011cascade} propose a boosting algorithm for jointly learning pruning and ranker stages.
\citet{culpepper2016dynamic} use a cascade of binary classifiers based on random forests; each classifier is used to predict whether to truncate the given ranked list at a specific cut-off value.
\citet{li2022certified} propose a score-thresholding method, which makes sure the trimmed candidate list produces re-ranking outcomes that satisfy the user-specified error tolerance of an \ac{IR} evaluation metric.
%
%Differing from the studies mentioned above, 
%Our focus is to reproduce \ac{RLT} methods for re-ranking instead of .
%We leave the exploration of candidate pruning methods~\citep{li2022certified, culpepper2016dynamic, wang2011cascade} can be used in our scenario
%, our target is to reproduce \ac{RLT} methods 

%\textit{Early stopping} is used for an on-the-fly re-ranker to stop scoring the rest of the items in a ranked list, to avoid traversing the entire ranked list.
%\citet{ying2020adaptive} determine when to stop an on-the-fly re-ranker by comparing the re-ranker scores and a score threshold.

%Although candidate pruning methods~\citep{li2022certified, culpepper2016dynamic, wang2011cascade} can be used in our scenario, previous studies study \ac{RLT} and candidate pruning separately and do not merge these two lines of research~\citep{bahri2023surprise,ma2022incorporating,wang2022mtcut,wu2021learning,bahri2020choppy,lien2019assumption}.
%
%Our focus in this reproducibility study is on specifically reproducing \ac{RLT} methods in a ``\textit{retrieve-then-re-rank}'' setup.
%
%Exploring all approaches that have the potential to trim the retrieved list and determining the optimal one is beyond the scope of our work.

We also differ from \citet{asadi2013effectiveness} and \citet{tonellotto2013efficient}, who investigate improving the efficiency of candidate generation, i.e., first-stage retrieval.
Specifically, \citet{tonellotto2013efficient} predict the number of candidate items that should be retrieved by the candidate generation algorithm WAND~\citep{broder2003efficient} on a per-query basis.
Our focus lies in improving re-ranking efficiency by truncating retrieved lists; in our setup, the retriever always returns a fixed number of items.

%% file: Sections/07_Conclusion.tex
\section{Conclusions \& Future Work}
We have reproduced numerous \ac{RLT} methods in the ``\textit{retrieve-then-re-rank}'' setup.
%We experimented with \ac{RLT} methods in three perspectives:
%\begin{enumerate*}[label=(\roman*)]
%    \item assessing \ac{RLT} methods that optimized to model different trade-offs between effectiveness and efficiency in re-ranking,
%    \item investigating the impact of different types of first-stage retrievers on \ac{RLT} methods, and 
%    \item investigating the impact of different types of re-rankers on \ac{RLT} methods.
%\end{enumerate*}
We showed that findings on \ac{RLT} do not generalize well to this new setup.
We found that
\begin{enumerate*}[label=(\roman*)]
\item supervised \ac{RLT} methods do not demonstrate a clear advantage over their unsupervised counterparts; potential fixed re-ranking depths can closely approximate the effectiveness/efficiency trade-off achieved by supervised methods;
%although supervised methods achieve better re-ranking effectiveness at lower cost compared to a fixed re-ranking depth of 1000 for pipelines with a BM25 retriever, other potential fixed-re-ranking depths can achieve very similar re-ranking effectiveness and efficiency;
%
\item distribution-based supervised methods achieve better effectiveness/efficiency trade-offs than their sequential labeling-based counterpart in most cases; the latter attains better re-ranking effectiveness at a lower cost for pipelines using BM25 retrieval;
%the sequential labeling-based \ac{RLT} method achieves better re-ranking effectiveness at a lower cost than their distribution-based supervised counterparts in the scenario prioritizing effectiveness and when BM25 as the retriever; however, distribution-based supervised methods show better effectiveness/efficiency trade-offs;
%
\item jointly learning \ac{RLT} with other tasks~\citep{wang2022mtcut} does not consistently yield a clear improvement; it only demonstrate a superior re-ranking effectiveness/efficiency trade-off for SPLADE++--RankLLaMA; and
%leads to a marked improvement in terms of re-ranking effectiveness/efficiency for the pipeline of SPLADE++--RankLLaMA, but does not benefit other combinations of other types of retrievers and re-rankers; and 
%
\item incorporating neural retriever embeddings~\citep{ma2022incorporating} does not exhibit a clear advantage; it merely yields marginal improvements in re-ranking effectiveness for RepLLaMA--RankLLaMA.
\end{enumerate*} 

We also learn valuable lessons from our experiments:
\begin{enumerate*}[label=(\roman*)]
\item the type of retriever significantly affects \ac{RLT} for re-ranking; with an effective retriever like SPLADE++ or RepLLaMA, a fixed re-ranking depth of 20 can already yield an excellent effectiveness/efficiency trade-off; increasing the fixed depth do not significantly improve effectiveness; using a fixed depth of 200 for retrieved lists returned by RepLLaMA, as done by \citet{ma2023fine}, results in unnecessary computational costs; and
\item the type of re-ranker (LLM or pre-trained LM-based) does not appear to influence the findings.
\end{enumerate*} 

We identify future directions: 
\begin{enumerate*}[label=(\roman*)]
\item the gap between Oracle and \ac{RLT} methods highlights the necessity of enhancing \ac{RLT} methods for re-ranking; we plan to solve the potential data scarcity issue highlighted in Section~\ref{sec:error}, and explore the use of \ac{QPP} methods for predicting query-specific re-ranking cut-offs~\citep{meng2024dc,meng2024query,arabzadeh2024query,meng2023query,meng2023Performance};
\item we only consider point-wise re-rankers; we plan to explore \ac{RLT} for pair-wise and list-wise \ac{LLM}-based re-rankers~\citep{qin2023large,zhang2023rank,pradeep2023rankzephyr,pradeep2023rankvicuna}; and
\item we plan to explore \ac{RLT} for re-ranking in \ac{CS}~\citep{abbasiantaeb2024llm,meng2023system,meng2021initiative,sun2021conversations,meng2020dukenet,meng2020refnet}.
\end{enumerate*}

%% file: Sections/08_Acknowledgement.tex
\subsubsection*{\bf Acknowledgments.}
We thank Yiding Liu (Baidu Inc.) for insightful discussions that contributed to the development of our ideas.
    This research was supported 
    by the China Scholarship Council (CSC) under grant number 202106220041,
    the Dutch Research Council (NWO), under project numbers 024.004.022, NWA.1389.20.\-183, and KICH3.LTP.20.006,
    and the 
    European Union's Horizon Europe program under grant agreement No 101070212.

%% file: main.bbl

\begin{thebibliography}{67}

%%% the Web address.

\ifx \showCODEN    \undefined \def \showCODEN     #1{\unskip}     \fi
\ifx \showDOI      \undefined \def \showDOI       #1{#1}\fi
\ifx \showISBNx    \undefined \def \showISBNx     #1{\unskip}     \fi
\ifx \showISBNxiii \undefined \def \showISBNxiii  #1{\unskip}     \fi
\ifx \showISSN     \undefined \def \showISSN      #1{\unskip}     \fi
\ifx \showLCCN     \undefined \def \showLCCN      #1{\unskip}     \fi
\ifx \shownote     \undefined \def \shownote      #1{#1}          \fi
\ifx \showarticletitle \undefined \def \showarticletitle #1{#1}   \fi
\ifx \showURL      \undefined \def \showURL       {\relax}        \fi
% invisible to TeX
\providecommand\bibfield[2]{#2}
\providecommand\bibinfo[2]{#2}
\providecommand\natexlab[1]{#1}
\providecommand\showeprint[2][]{arXiv:#2}

\bibitem[Abbasiantaeb et~al\mbox{.}(2023)]%
        {abbasiantaeb2024llm}
\bibfield{author}{\bibinfo{person}{Zahra Abbasiantaeb}, \bibinfo{person}{Chuan Meng}, \bibinfo{person}{David Rau}, \bibinfo{person}{Antonis Krasakis}, \bibinfo{person}{Hossein~A Rahmani}, {and} \bibinfo{person}{Mohammad Aliannejadi}.} \bibinfo{year}{2023}\natexlab{}.
\newblock \showarticletitle{LLM-based Retrieval and Generation Pipelines for TREC Interactive Knowledge Assistance Track (iKAT) 2023}. In \bibinfo{booktitle}{\emph{TREC}}.
\newblock


\bibitem[Arabzadeh et~al\mbox{.}(2024)]%
        {arabzadeh2024query}
\bibfield{author}{\bibinfo{person}{Negar Arabzadeh}, \bibinfo{person}{Chuan Meng}, \bibinfo{person}{Mohammad Aliannejadi}, {and} \bibinfo{person}{Ebrahim Bagheri}.} \bibinfo{year}{2024}\natexlab{}.
\newblock \showarticletitle{Query Performance Prediction: From Fundamentals to Advanced Techniques}. In \bibinfo{booktitle}{\emph{ECIR}}. Springer, \bibinfo{pages}{381--388}.
\newblock


\bibitem[Arampatzis et~al\mbox{.}(2009)]%
        {arampatzis2009stop}
\bibfield{author}{\bibinfo{person}{Avi Arampatzis}, \bibinfo{person}{Jaap Kamps}, {and} \bibinfo{person}{Stephen Robertson}.} \bibinfo{year}{2009}\natexlab{}.
\newblock \showarticletitle{Where to Stop Reading a Ranked List? Threshold Optimization using Truncated Score Distributions}. In \bibinfo{booktitle}{\emph{SIGIR}}. \bibinfo{pages}{524--531}.
\newblock


\bibitem[Asadi and Lin(2013)]%
        {asadi2013effectiveness}
\bibfield{author}{\bibinfo{person}{Nima Asadi} {and} \bibinfo{person}{Jimmy Lin}.} \bibinfo{year}{2013}\natexlab{}.
\newblock \showarticletitle{Effectiveness/Efficiency Tradeoffs for Candidate Generation in Multi-Stage Retrieval Architectures}. In \bibinfo{booktitle}{\emph{SIGIR}}. \bibinfo{pages}{997--1000}.
\newblock


\bibitem[Askari et~al\mbox{.}(2024)]%
        {askari2024self}
\bibfield{author}{\bibinfo{person}{Arian Askari}, \bibinfo{person}{Roxana Petcu}, \bibinfo{person}{Chuan Meng}, \bibinfo{person}{Mohammad Aliannejadi}, \bibinfo{person}{Amin Abolghasemi}, \bibinfo{person}{Evangelos Kanoulas}, {and} \bibinfo{person}{Suzan Verberne}.} \bibinfo{year}{2024}\natexlab{}.
\newblock \showarticletitle{Self-seeding and Multi-intent Self-instructing LLMs for Generating Intent-aware Information-Seeking dialogs}.
\newblock \bibinfo{journal}{\emph{arXiv preprint arXiv:2402.11633}} (\bibinfo{year}{2024}).
\newblock


\bibitem[Bahri et~al\mbox{.}(2020)]%
        {bahri2020choppy}
\bibfield{author}{\bibinfo{person}{Dara Bahri}, \bibinfo{person}{Yi Tay}, \bibinfo{person}{Che Zheng}, \bibinfo{person}{Donald Metzler}, {and} \bibinfo{person}{Andrew Tomkins}.} \bibinfo{year}{2020}\natexlab{}.
\newblock \showarticletitle{Choppy: Cut Transformer for Ranked List Truncation}. In \bibinfo{booktitle}{\emph{SIGIR}}. \bibinfo{pages}{1513--1516}.
\newblock


\bibitem[Bahri et~al\mbox{.}(2023)]%
        {bahri2023surprise}
\bibfield{author}{\bibinfo{person}{Dara Bahri}, \bibinfo{person}{Che Zheng}, \bibinfo{person}{Yi Tay}, \bibinfo{person}{Donald Metzler}, {and} \bibinfo{person}{Andrew Tomkins}.} \bibinfo{year}{2023}\natexlab{}.
\newblock \showarticletitle{Surprise: Result List Truncation via Extreme Value Theory}. In \bibinfo{booktitle}{\emph{SIGIR}}. \bibinfo{pages}{2404--2408}.
\newblock


\bibitem[Broder et~al\mbox{.}(2003)]%
        {broder2003efficient}
\bibfield{author}{\bibinfo{person}{Andrei~Z Broder}, \bibinfo{person}{David Carmel}, \bibinfo{person}{Michael Herscovici}, \bibinfo{person}{Aya Soffer}, {and} \bibinfo{person}{Jason Zien}.} \bibinfo{year}{2003}\natexlab{}.
\newblock \showarticletitle{Efficient Query Evaluation using a Two-Level Retrieval Process}. In \bibinfo{booktitle}{\emph{CIKM}}. \bibinfo{pages}{426--434}.
\newblock


\bibitem[Bruch et~al\mbox{.}(2023)]%
        {bruch2023efficient}
\bibfield{author}{\bibinfo{person}{Sebastian Bruch}, \bibinfo{person}{Claudio Lucchese}, {and} \bibinfo{person}{Franco~Maria Nardini}.} \bibinfo{year}{2023}\natexlab{}.
\newblock \showarticletitle{Efficient and Effective Tree-based and Neural Learning to Rank}.
\newblock \bibinfo{journal}{\emph{Foundations and Trends in Information Retrieval}} \bibinfo{volume}{17}, \bibinfo{number}{1} (\bibinfo{year}{2023}), \bibinfo{pages}{1--123}.
\newblock


\bibitem[Cohen et~al\mbox{.}(2021)]%
        {cohen2021not}
\bibfield{author}{\bibinfo{person}{Daniel Cohen}, \bibinfo{person}{Bhaskar Mitra}, \bibinfo{person}{Oleg Lesota}, \bibinfo{person}{Navid Rekabsaz}, {and} \bibinfo{person}{Carsten Eickhoff}.} \bibinfo{year}{2021}\natexlab{}.
\newblock \showarticletitle{Not All Relevance Scores are Equal: Efficient Uncertainty and Calibration Modeling for Deep Retrieval Models}. In \bibinfo{booktitle}{\emph{SIGIR}}. \bibinfo{pages}{654--664}.
\newblock


\bibitem[Craswell et~al\mbox{.}(2020)]%
        {craswell2020}
\bibfield{author}{\bibinfo{person}{Nick Craswell}, \bibinfo{person}{Bhaskar Mitra}, \bibinfo{person}{Emine Yilmaz}, {and} \bibinfo{person}{Daniel Campos}.} \bibinfo{year}{2020}\natexlab{}.
\newblock \showarticletitle{Overview of the TREC 2020 Deep Learning Track}. In \bibinfo{booktitle}{\emph{TREC}}.
\newblock


\bibitem[Craswell et~al\mbox{.}(2019)]%
        {craswell2019}
\bibfield{author}{\bibinfo{person}{Nick Craswell}, \bibinfo{person}{Bhaskar Mitra}, \bibinfo{person}{Emine Yilmaz}, \bibinfo{person}{Daniel Campos}, {and} \bibinfo{person}{Ellen~M Voorhees}.} \bibinfo{year}{2019}\natexlab{}.
\newblock \showarticletitle{Overview of the TREC 2019 Deep Learning Track}. In \bibinfo{booktitle}{\emph{TREC}}.
\newblock


\bibitem[Culpepper et~al\mbox{.}(2016)]%
        {culpepper2016dynamic}
\bibfield{author}{\bibinfo{person}{J~Shane Culpepper}, \bibinfo{person}{Charles~LA Clarke}, {and} \bibinfo{person}{Jimmy Lin}.} \bibinfo{year}{2016}\natexlab{}.
\newblock \showarticletitle{Dynamic Cutoff Prediction in Multi-Stage Retrieval Systems}. In \bibinfo{booktitle}{\emph{Proceedings of the 21st Australasian Document Computing Symposium}}. \bibinfo{pages}{17--24}.
\newblock


\bibitem[Darling(1957)]%
        {darling1957kolmogorov}
\bibfield{author}{\bibinfo{person}{Donald~A Darling}.} \bibinfo{year}{1957}\natexlab{}.
\newblock \showarticletitle{The Kolmogorov-Smirnov, Cramer-Von Mises Tests}.
\newblock \bibinfo{journal}{\emph{The Annals of Mathematical Statistics}} \bibinfo{volume}{28}, \bibinfo{number}{4} (\bibinfo{year}{1957}), \bibinfo{pages}{823--838}.
\newblock


\bibitem[Devlin et~al\mbox{.}(2019)]%
        {devlin2019bert}
\bibfield{author}{\bibinfo{person}{Jacob Devlin}, \bibinfo{person}{Ming-Wei Chang}, \bibinfo{person}{Kenton Lee}, {and} \bibinfo{person}{Kristina Toutanova}.} \bibinfo{year}{2019}\natexlab{}.
\newblock \showarticletitle{BERT: Pre-training of Deep Bidirectional Transformers for Language Understanding}. In \bibinfo{booktitle}{\emph{NAACL}}. \bibinfo{pages}{4171--4186}.
\newblock


\bibitem[Drozdov et~al\mbox{.}(2023)]%
        {drozdov2023parade}
\bibfield{author}{\bibinfo{person}{Andrew Drozdov}, \bibinfo{person}{Honglei Zhuang}, \bibinfo{person}{Zhuyun Dai}, \bibinfo{person}{Zhen Qin}, \bibinfo{person}{Razieh Rahimi}, \bibinfo{person}{Xuanhui Wang}, \bibinfo{person}{Dana Alon}, \bibinfo{person}{Mohit Iyyer}, \bibinfo{person}{Andrew McCallum}, \bibinfo{person}{Donald Metzler}, {et~al\mbox{.}}} \bibinfo{year}{2023}\natexlab{}.
\newblock \showarticletitle{{P}a{R}a{D}e: Passage Ranking using Demonstrations with {LLM}s}. In \bibinfo{booktitle}{\emph{Findings of EMNLP}}. \bibinfo{pages}{14242--14252}.
\newblock


\bibitem[Formal et~al\mbox{.}(2022)]%
        {formal2022distillation}
\bibfield{author}{\bibinfo{person}{Thibault Formal}, \bibinfo{person}{Carlos Lassance}, \bibinfo{person}{Benjamin Piwowarski}, {and} \bibinfo{person}{St{\'e}phane Clinchant}.} \bibinfo{year}{2022}\natexlab{}.
\newblock \showarticletitle{From Distillation to Hard Negative Sampling: Making Sparse Neural IR Models More Effective}. In \bibinfo{booktitle}{\emph{SIGIR}}. \bibinfo{pages}{2353--2359}.
\newblock


\bibitem[Gao et~al\mbox{.}(2020)]%
        {gao2020understanding}
\bibfield{author}{\bibinfo{person}{Luyu Gao}, \bibinfo{person}{Zhuyun Dai}, {and} \bibinfo{person}{Jamie Callan}.} \bibinfo{year}{2020}\natexlab{}.
\newblock \showarticletitle{Understanding BERT Rankers Under Distillation}. In \bibinfo{booktitle}{\emph{SIGIR}}. \bibinfo{pages}{149--152}.
\newblock


\bibitem[Gienapp et~al\mbox{.}(2022)]%
        {gienapp2022sparse}
\bibfield{author}{\bibinfo{person}{Lukas Gienapp}, \bibinfo{person}{Maik Fr{\"o}be}, \bibinfo{person}{Matthias Hagen}, {and} \bibinfo{person}{Martin Potthast}.} \bibinfo{year}{2022}\natexlab{}.
\newblock \showarticletitle{Sparse Pairwise Re-ranking with Pre-trained Transformers}. In \bibinfo{booktitle}{\emph{ICTIR}}. \bibinfo{pages}{72--80}.
\newblock


\bibitem[Hofst{\"a}tter et~al\mbox{.}(2020)]%
        {hofstatter2020interpretable}
\bibfield{author}{\bibinfo{person}{Sebastian Hofst{\"a}tter}, \bibinfo{person}{Markus Zlabinger}, {and} \bibinfo{person}{Allan Hanbury}.} \bibinfo{year}{2020}\natexlab{}.
\newblock \showarticletitle{Interpretable \& Time-Budget-Constrained Contextualization for Re-Ranking}. In \bibinfo{booktitle}{\emph{ECAI 2020 24th European Conference on Artificial Intelligence, 29 August-8 September 2020, Santiago de Compostela, Spain-Including 10th Conference on Prestigious Applications of Artificial Intelligence (PAIS 2020)}}. IOS Press, \bibinfo{pages}{1--8}.
\newblock


\bibitem[Kingma and Ba(2015)]%
        {kingma2014adam}
\bibfield{author}{\bibinfo{person}{Diederik~P Kingma} {and} \bibinfo{person}{Jimmy Ba}.} \bibinfo{year}{2015}\natexlab{}.
\newblock \showarticletitle{Adam: A Method for Stochastic Optimization}. In \bibinfo{booktitle}{\emph{ICLR}}.
\newblock


\bibitem[Le and Mikolov(2014)]%
        {le2014distributed}
\bibfield{author}{\bibinfo{person}{Quoc Le} {and} \bibinfo{person}{Tomas Mikolov}.} \bibinfo{year}{2014}\natexlab{}.
\newblock \showarticletitle{Distributed Representations of Sentences and Documents}. In \bibinfo{booktitle}{\emph{ICML}}. PMLR, \bibinfo{pages}{1188--1196}.
\newblock


\bibitem[Lesota et~al\mbox{.}(2021)]%
        {lesota2021modern}
\bibfield{author}{\bibinfo{person}{Oleg Lesota}, \bibinfo{person}{Navid Rekabsaz}, \bibinfo{person}{Daniel Cohen}, \bibinfo{person}{Klaus~Antonius Grasserbauer}, \bibinfo{person}{Carsten Eickhoff}, {and} \bibinfo{person}{Markus Schedl}.} \bibinfo{year}{2021}\natexlab{}.
\newblock \showarticletitle{A Modern Perspective on Query Likelihood with Deep Generative Retrieval Models}. In \bibinfo{booktitle}{\emph{ICTIR}}. \bibinfo{pages}{185--195}.
\newblock


\bibitem[Li et~al\mbox{.}(2022)]%
        {li2022certified}
\bibfield{author}{\bibinfo{person}{Minghan Li}, \bibinfo{person}{Xinyu Zhang}, \bibinfo{person}{Ji Xin}, \bibinfo{person}{Hongyang Zhang}, {and} \bibinfo{person}{Jimmy Lin}.} \bibinfo{year}{2022}\natexlab{}.
\newblock \showarticletitle{Certified Error Control of Candidate Set Pruning for Two-Stage Relevance Ranking}. In \bibinfo{booktitle}{\emph{EMNLP}}. \bibinfo{pages}{333--345}.
\newblock


\bibitem[Lien et~al\mbox{.}(2019)]%
        {lien2019assumption}
\bibfield{author}{\bibinfo{person}{Yen-Chieh Lien}, \bibinfo{person}{Daniel Cohen}, {and} \bibinfo{person}{W~Bruce Croft}.} \bibinfo{year}{2019}\natexlab{}.
\newblock \showarticletitle{An Assumption-Free Approach to the Dynamic Truncation of Ranked Lists}. In \bibinfo{booktitle}{\emph{ICTIR}}. \bibinfo{pages}{79--82}.
\newblock


\bibitem[Lupu and Hanbury(2013)]%
        {lupu2013patent}
\bibfield{author}{\bibinfo{person}{Mihai Lupu} {and} \bibinfo{person}{Allan Hanbury}.} \bibinfo{year}{2013}\natexlab{}.
\newblock \showarticletitle{Patent Retrieval}.
\newblock \bibinfo{journal}{\emph{Foundations and Trends in Information Retrieval}} \bibinfo{volume}{7}, \bibinfo{number}{1} (\bibinfo{year}{2013}), \bibinfo{pages}{1--97}.
\newblock


\bibitem[Ma et~al\mbox{.}(2023a)]%
        {ma2023fine}
\bibfield{author}{\bibinfo{person}{Xueguang Ma}, \bibinfo{person}{Liang Wang}, \bibinfo{person}{Nan Yang}, \bibinfo{person}{Furu Wei}, {and} \bibinfo{person}{Jimmy Lin}.} \bibinfo{year}{2023}\natexlab{a}.
\newblock \showarticletitle{Fine-Tuning LLaMA for Multi-Stage Text Retrieval}.
\newblock \bibinfo{journal}{\emph{arXiv preprint arXiv:2310.08319}} (\bibinfo{year}{2023}).
\newblock


\bibitem[Ma et~al\mbox{.}(2023b)]%
        {ma2023zero}
\bibfield{author}{\bibinfo{person}{Xueguang Ma}, \bibinfo{person}{Xinyu Zhang}, \bibinfo{person}{Ronak Pradeep}, {and} \bibinfo{person}{Jimmy Lin}.} \bibinfo{year}{2023}\natexlab{b}.
\newblock \showarticletitle{Zero-Shot Listwise Document Reranking with a Large Language Model}.
\newblock \bibinfo{journal}{\emph{arXiv preprint arXiv:2305.02156}} (\bibinfo{year}{2023}).
\newblock


\bibitem[Ma et~al\mbox{.}(2022)]%
        {ma2022incorporating}
\bibfield{author}{\bibinfo{person}{Yixiao Ma}, \bibinfo{person}{Qingyao Ai}, \bibinfo{person}{Yueyue Wu}, \bibinfo{person}{Yunqiu Shao}, \bibinfo{person}{Yiqun Liu}, \bibinfo{person}{Min Zhang}, {and} \bibinfo{person}{Shaoping Ma}.} \bibinfo{year}{2022}\natexlab{}.
\newblock \showarticletitle{Incorporating Retrieval Information into the Truncation of Ranking Lists for Better Legal Search}. In \bibinfo{booktitle}{\emph{SIGIR}}. \bibinfo{pages}{438--448}.
\newblock


\bibitem[MacAvaney et~al\mbox{.}(2020)]%
        {macavaney2020efficient}
\bibfield{author}{\bibinfo{person}{Sean MacAvaney}, \bibinfo{person}{Franco~Maria Nardini}, \bibinfo{person}{Raffaele Perego}, \bibinfo{person}{Nicola Tonellotto}, \bibinfo{person}{Nazli Goharian}, {and} \bibinfo{person}{Ophir Frieder}.} \bibinfo{year}{2020}\natexlab{}.
\newblock \showarticletitle{Efficient Document Re-Ranking for Transformers by Precomputing Term Representations}. In \bibinfo{booktitle}{\emph{SIGIR}}. \bibinfo{pages}{49--58}.
\newblock


\bibitem[MacAvaney et~al\mbox{.}(2022)]%
        {macavaney2022adaptive}
\bibfield{author}{\bibinfo{person}{Sean MacAvaney}, \bibinfo{person}{Nicola Tonellotto}, {and} \bibinfo{person}{Craig Macdonald}.} \bibinfo{year}{2022}\natexlab{}.
\newblock \showarticletitle{Adaptive Re-Ranking with a Corpus Graph}. In \bibinfo{booktitle}{\emph{CIKM}}. \bibinfo{pages}{1491--1500}.
\newblock


\bibitem[Manmatha et~al\mbox{.}(2001)]%
        {manmatha2001modeling}
\bibfield{author}{\bibinfo{person}{Raghavan Manmatha}, \bibinfo{person}{Toni Rath}, {and} \bibinfo{person}{Fangfang Feng}.} \bibinfo{year}{2001}\natexlab{}.
\newblock \showarticletitle{Modeling Score Distributions for Combining the Outputs of Search Engines}. In \bibinfo{booktitle}{\emph{SIGIR}}. \bibinfo{pages}{267--275}.
\newblock


\bibitem[Matsubara et~al\mbox{.}(2020)]%
        {matsubara2020reranking}
\bibfield{author}{\bibinfo{person}{Yoshitomo Matsubara}, \bibinfo{person}{Thuy Vu}, {and} \bibinfo{person}{Alessandro Moschitti}.} \bibinfo{year}{2020}\natexlab{}.
\newblock \showarticletitle{Reranking for Efficient Transformer-based Answer Selection}. In \bibinfo{booktitle}{\emph{SIGIR}}. \bibinfo{pages}{1577--1580}.
\newblock


\bibitem[Meng(2024)]%
        {meng2024dc}
\bibfield{author}{\bibinfo{person}{Chuan Meng}.} \bibinfo{year}{2024}\natexlab{}.
\newblock \showarticletitle{Query Performance Prediction for Conversational Search and Beyond}. In \bibinfo{booktitle}{\emph{SIGIR}}.
\newblock


\bibitem[Meng et~al\mbox{.}(2023a)]%
        {meng2023Performance}
\bibfield{author}{\bibinfo{person}{Chuan Meng}, \bibinfo{person}{Mohammad Aliannejadi}, {and} \bibinfo{person}{Maarten de Rijke}.} \bibinfo{year}{2023}\natexlab{a}.
\newblock \showarticletitle{Performance Prediction for Conversational Search Using Perplexities of Query Rewrites}. In \bibinfo{booktitle}{\emph{QPP++2023}}. \bibinfo{pages}{25--28}.
\newblock


\bibitem[Meng et~al\mbox{.}(2023b)]%
        {meng2023system}
\bibfield{author}{\bibinfo{person}{Chuan Meng}, \bibinfo{person}{Mohammad Aliannejadi}, {and} \bibinfo{person}{Maarten de Rijke}.} \bibinfo{year}{2023}\natexlab{b}.
\newblock \showarticletitle{System Initiative Prediction for Multi-turn Conversational Information Seeking}. In \bibinfo{booktitle}{\emph{CIKM}}. \bibinfo{pages}{1807--1817}.
\newblock


\bibitem[Meng et~al\mbox{.}(2023c)]%
        {meng2023query}
\bibfield{author}{\bibinfo{person}{Chuan Meng}, \bibinfo{person}{Negar Arabzadeh}, \bibinfo{person}{Mohammad Aliannejadi}, {and} \bibinfo{person}{Maarten de Rijke}.} \bibinfo{year}{2023}\natexlab{c}.
\newblock \showarticletitle{Query Performance Prediction: From Ad-hoc to Conversational Search}. In \bibinfo{booktitle}{\emph{SIGIR}}. \bibinfo{pages}{2583–2593}.
\newblock


\bibitem[Meng et~al\mbox{.}(2024)]%
        {meng2024query}
\bibfield{author}{\bibinfo{person}{Chuan Meng}, \bibinfo{person}{Negar Arabzadeh}, \bibinfo{person}{Arian Askari}, \bibinfo{person}{Mohammad Aliannejadi}, {and} \bibinfo{person}{Maarten de Rijke}.} \bibinfo{year}{2024}\natexlab{}.
\newblock \showarticletitle{Query Performance Prediction using Relevance Judgments Generated by Large Language Models}.
\newblock \bibinfo{journal}{\emph{arXiv preprint arXiv:2404.01012}} (\bibinfo{year}{2024}).
\newblock


\bibitem[Meng et~al\mbox{.}(2020a)]%
        {meng2020refnet}
\bibfield{author}{\bibinfo{person}{Chuan Meng}, \bibinfo{person}{Pengjie Ren}, \bibinfo{person}{Zhumin Chen}, \bibinfo{person}{Christof Monz}, \bibinfo{person}{Jun Ma}, {and} \bibinfo{person}{Maarten de Rijke}.} \bibinfo{year}{2020}\natexlab{a}.
\newblock \showarticletitle{RefNet: A Reference-aware Network for Background Based Conversation}. In \bibinfo{booktitle}{\emph{AAAI}}.
\newblock


\bibitem[Meng et~al\mbox{.}(2021)]%
        {meng2021initiative}
\bibfield{author}{\bibinfo{person}{Chuan Meng}, \bibinfo{person}{Pengjie Ren}, \bibinfo{person}{Zhumin Chen}, \bibinfo{person}{Zhaochun Ren}, \bibinfo{person}{Tengxiao Xi}, {and} \bibinfo{person}{Maarten~de Rijke}.} \bibinfo{year}{2021}\natexlab{}.
\newblock \showarticletitle{Initiative-Aware Self-Supervised Learning for Knowledge-Grounded Conversations}. In \bibinfo{booktitle}{\emph{SIGIR}}. \bibinfo{pages}{522–532}.
\newblock


\bibitem[Meng et~al\mbox{.}(2020b)]%
        {meng2020dukenet}
\bibfield{author}{\bibinfo{person}{Chuan Meng}, \bibinfo{person}{Pengjie Ren}, \bibinfo{person}{Zhumin Chen}, \bibinfo{person}{Weiwei Sun}, \bibinfo{person}{Zhaochun Ren}, \bibinfo{person}{Zhaopeng Tu}, {and} \bibinfo{person}{Maarten~de Rijke}.} \bibinfo{year}{2020}\natexlab{b}.
\newblock \showarticletitle{DukeNet: A Dual Knowledge Interaction Network for Knowledge-Grounded Conversation}. In \bibinfo{booktitle}{\emph{SIGIR}}. \bibinfo{pages}{1151–1160}.
\newblock


\bibitem[Nogueira and Cho(2019)]%
        {nogueira2019passage}
\bibfield{author}{\bibinfo{person}{Rodrigo Nogueira} {and} \bibinfo{person}{Kyunghyun Cho}.} \bibinfo{year}{2019}\natexlab{}.
\newblock \showarticletitle{Passage Re-ranking with BERT}.
\newblock \bibinfo{journal}{\emph{arXiv preprint arXiv:1901.04085}} (\bibinfo{year}{2019}).
\newblock


\bibitem[Nogueira et~al\mbox{.}(2020)]%
        {nogueira2020document}
\bibfield{author}{\bibinfo{person}{Rodrigo Nogueira}, \bibinfo{person}{Zhiying Jiang}, \bibinfo{person}{Ronak Pradeep}, {and} \bibinfo{person}{Jimmy Lin}.} \bibinfo{year}{2020}\natexlab{}.
\newblock \showarticletitle{Document Ranking with a Pretrained Sequence-to-Sequence Model}. In \bibinfo{booktitle}{\emph{EMNLP}}. \bibinfo{pages}{708--718}.
\newblock


\bibitem[Nogueira et~al\mbox{.}(2019)]%
        {nogueira2019multi}
\bibfield{author}{\bibinfo{person}{Rodrigo Nogueira}, \bibinfo{person}{Wei Yang}, \bibinfo{person}{Kyunghyun Cho}, {and} \bibinfo{person}{Jimmy Lin}.} \bibinfo{year}{2019}\natexlab{}.
\newblock \showarticletitle{Multi-Stage Document Ranking with BERT}.
\newblock \bibinfo{journal}{\emph{arXiv preprint arXiv:1910.14424}} (\bibinfo{year}{2019}).
\newblock


\bibitem[Pickands~III(1975)]%
        {pickands1975statistical}
\bibfield{author}{\bibinfo{person}{James Pickands~III}.} \bibinfo{year}{1975}\natexlab{}.
\newblock \showarticletitle{Statistical Inference Using Extreme Order Statistics}.
\newblock \bibinfo{journal}{\emph{the Annals of Statistics}} (\bibinfo{year}{1975}), \bibinfo{pages}{119--131}.
\newblock


\bibitem[Pradeep et~al\mbox{.}(2023a)]%
        {pradeep2023rankvicuna}
\bibfield{author}{\bibinfo{person}{Ronak Pradeep}, \bibinfo{person}{Sahel Sharifymoghaddam}, {and} \bibinfo{person}{Jimmy Lin}.} \bibinfo{year}{2023}\natexlab{a}.
\newblock \showarticletitle{RankVicuna: Zero-Shot Listwise Document Reranking with Open-Source Large Language Models}.
\newblock \bibinfo{journal}{\emph{arXiv preprint arXiv:2309.15088}} (\bibinfo{year}{2023}).
\newblock


\bibitem[Pradeep et~al\mbox{.}(2023b)]%
        {pradeep2023rankzephyr}
\bibfield{author}{\bibinfo{person}{Ronak Pradeep}, \bibinfo{person}{Sahel Sharifymoghaddam}, {and} \bibinfo{person}{Jimmy Lin}.} \bibinfo{year}{2023}\natexlab{b}.
\newblock \showarticletitle{RankZephyr: Effective and Robust Zero-Shot Listwise Reranking is a Breeze!}
\newblock \bibinfo{journal}{\emph{arXiv preprint arXiv:2312.02724}} (\bibinfo{year}{2023}).
\newblock


\bibitem[Qin et~al\mbox{.}(2023)]%
        {qin2023large}
\bibfield{author}{\bibinfo{person}{Zhen Qin}, \bibinfo{person}{Rolf Jagerman}, \bibinfo{person}{Kai Hui}, \bibinfo{person}{Honglei Zhuang}, \bibinfo{person}{Junru Wu}, \bibinfo{person}{Jiaming Shen}, \bibinfo{person}{Tianqi Liu}, \bibinfo{person}{Jialu Liu}, \bibinfo{person}{Donald Metzler}, \bibinfo{person}{Xuanhui Wang}, {et~al\mbox{.}}} \bibinfo{year}{2023}\natexlab{}.
\newblock \showarticletitle{Large Language Models are Effective Text Rankers with Pairwise Ranking Prompting}.
\newblock \bibinfo{journal}{\emph{arXiv preprint arXiv:2306.17563}} (\bibinfo{year}{2023}).
\newblock


\bibitem[Robertson and Zaragoza(2009)]%
        {robertson2009probabilistic}
\bibfield{author}{\bibinfo{person}{Stephen Robertson} {and} \bibinfo{person}{Hugo Zaragoza}.} \bibinfo{year}{2009}\natexlab{}.
\newblock \showarticletitle{The Probabilistic Relevance Framework: BM25 and Beyond}.
\newblock \bibinfo{journal}{\emph{Foundations and Trends in Information Retrieval}} \bibinfo{volume}{3}, \bibinfo{number}{4} (\bibinfo{year}{2009}), \bibinfo{pages}{333--389}.
\newblock


\bibitem[Sachan et~al\mbox{.}(2022)]%
        {sachan2022improving}
\bibfield{author}{\bibinfo{person}{Devendra Sachan}, \bibinfo{person}{Mike Lewis}, \bibinfo{person}{Mandar Joshi}, \bibinfo{person}{Armen Aghajanyan}, \bibinfo{person}{Wen-tau Yih}, \bibinfo{person}{Joelle Pineau}, {and} \bibinfo{person}{Luke Zettlemoyer}.} \bibinfo{year}{2022}\natexlab{}.
\newblock \showarticletitle{Improving Passage Retrieval with Zero-Shot Question Generation}. In \bibinfo{booktitle}{\emph{EMNLP}}. \bibinfo{pages}{3781--3797}.
\newblock


\bibitem[Soldaini and Moschitti(2020)]%
        {soldaini2020cascade}
\bibfield{author}{\bibinfo{person}{Luca Soldaini} {and} \bibinfo{person}{Alessandro Moschitti}.} \bibinfo{year}{2020}\natexlab{}.
\newblock \showarticletitle{The Cascade Transformer: an Application for Efficient Answer Sentence Selection}. In \bibinfo{booktitle}{\emph{Proceedings of the 58th Annual Meeting of the Association for Computational Linguistics}}. \bibinfo{pages}{5697--5708}.
\newblock


\bibitem[Sun et~al\mbox{.}(2021)]%
        {sun2021conversations}
\bibfield{author}{\bibinfo{person}{Weiwei Sun}, \bibinfo{person}{Chuan Meng}, \bibinfo{person}{Qi Meng}, \bibinfo{person}{Zhaochun Ren}, \bibinfo{person}{Pengjie Ren}, \bibinfo{person}{Zhumin Chen}, {and} \bibinfo{person}{Maarten~de Rijke}.} \bibinfo{year}{2021}\natexlab{}.
\newblock \showarticletitle{Conversations Powered by Cross-Lingual Knowledge}. In \bibinfo{booktitle}{\emph{SIGIR}}. \bibinfo{pages}{1442--1451}.
\newblock


\bibitem[Sun et~al\mbox{.}(2023)]%
        {sun2023chatgpt}
\bibfield{author}{\bibinfo{person}{Weiwei Sun}, \bibinfo{person}{Lingyong Yan}, \bibinfo{person}{Xinyu Ma}, \bibinfo{person}{Pengjie Ren}, \bibinfo{person}{Dawei Yin}, {and} \bibinfo{person}{Zhaochun Ren}.} \bibinfo{year}{2023}\natexlab{}.
\newblock \showarticletitle{Is ChatGPT Good at Search? Investigating Large Language Models as Re-Ranking Agent}. In \bibinfo{booktitle}{\emph{EMNLP}}. \bibinfo{pages}{14918--14937}.
\newblock


\bibitem[Tomlinson et~al\mbox{.}(2007)]%
        {tomlinson2007overview}
\bibfield{author}{\bibinfo{person}{Stephen Tomlinson}, \bibinfo{person}{Douglas~W Oard}, \bibinfo{person}{Jason~R Baron}, {and} \bibinfo{person}{Paul Thompson}.} \bibinfo{year}{2007}\natexlab{}.
\newblock \showarticletitle{Overview of the TREC 2007 Legal Track.}. In \bibinfo{booktitle}{\emph{TREC}}.
\newblock


\bibitem[Tonellotto et~al\mbox{.}(2013)]%
        {tonellotto2013efficient}
\bibfield{author}{\bibinfo{person}{Nicola Tonellotto}, \bibinfo{person}{Craig Macdonald}, {and} \bibinfo{person}{Iadh Ounis}.} \bibinfo{year}{2013}\natexlab{}.
\newblock \showarticletitle{Efficient and Effective Retrieval using Selective Pruning}. In \bibinfo{booktitle}{\emph{WSDM}}. \bibinfo{pages}{63--72}.
\newblock


\bibitem[Vaswani et~al\mbox{.}(2017)]%
        {vaswani2017attention}
\bibfield{author}{\bibinfo{person}{Ashish Vaswani}, \bibinfo{person}{Noam Shazeer}, \bibinfo{person}{Niki Parmar}, \bibinfo{person}{Jakob Uszkoreit}, \bibinfo{person}{Llion Jones}, \bibinfo{person}{Aidan~N Gomez}, \bibinfo{person}{{\L}ukasz Kaiser}, {and} \bibinfo{person}{Illia Polosukhin}.} \bibinfo{year}{2017}\natexlab{}.
\newblock \showarticletitle{Attention Is All You Need}. In \bibinfo{booktitle}{\emph{NeurIPS}}. \bibinfo{pages}{5998--6008}.
\newblock


\bibitem[Wang et~al\mbox{.}(2022)]%
        {wang2022mtcut}
\bibfield{author}{\bibinfo{person}{Dong Wang}, \bibinfo{person}{Jianxin Li}, \bibinfo{person}{Tianchen Zhu}, \bibinfo{person}{Haoyi Zhou}, \bibinfo{person}{Qishan Zhu}, \bibinfo{person}{Yuxin Wen}, {and} \bibinfo{person}{Hongming Piao}.} \bibinfo{year}{2022}\natexlab{}.
\newblock \showarticletitle{MtCut: A Multi-Task Framework for Ranked List Truncation}. In \bibinfo{booktitle}{\emph{WSDM}}. \bibinfo{pages}{1054--1062}.
\newblock


\bibitem[Wang et~al\mbox{.}(2010)]%
        {wang2010learning}
\bibfield{author}{\bibinfo{person}{Lidan Wang}, \bibinfo{person}{Jimmy Lin}, {and} \bibinfo{person}{Donald Metzler}.} \bibinfo{year}{2010}\natexlab{}.
\newblock \showarticletitle{Learning to Efficiently Rank}. In \bibinfo{booktitle}{\emph{SIGIR}}. \bibinfo{pages}{138--145}.
\newblock


\bibitem[Wang et~al\mbox{.}(2011)]%
        {wang2011cascade}
\bibfield{author}{\bibinfo{person}{Lidan Wang}, \bibinfo{person}{Jimmy Lin}, {and} \bibinfo{person}{Donald Metzler}.} \bibinfo{year}{2011}\natexlab{}.
\newblock \showarticletitle{A Cascade Ranking Model for Efficient Ranked Retrieval}. In \bibinfo{booktitle}{\emph{SIGIR}}. \bibinfo{pages}{105--114}.
\newblock


\bibitem[Wu et~al\mbox{.}(2021)]%
        {wu2021learning}
\bibfield{author}{\bibinfo{person}{Chen Wu}, \bibinfo{person}{Ruqing Zhang}, \bibinfo{person}{Jiafeng Guo}, \bibinfo{person}{Yixing Fan}, \bibinfo{person}{Yanyan Lan}, {and} \bibinfo{person}{Xueqi Cheng}.} \bibinfo{year}{2021}\natexlab{}.
\newblock \showarticletitle{Learning to Truncate Ranked Lists for Information Retrieval}. In \bibinfo{booktitle}{\emph{AAAI}}, Vol.~\bibinfo{volume}{35}. \bibinfo{pages}{4453--4461}.
\newblock


\bibitem[Xin et~al\mbox{.}(2020)]%
        {xin2020early}
\bibfield{author}{\bibinfo{person}{Ji Xin}, \bibinfo{person}{Rodrigo Nogueira}, \bibinfo{person}{Yaoliang Yu}, {and} \bibinfo{person}{Jimmy Lin}.} \bibinfo{year}{2020}\natexlab{}.
\newblock \showarticletitle{Early Exiting BERT for Efficient Document Ranking}. In \bibinfo{booktitle}{\emph{Proceedings of SustaiNLP: Workshop on Simple and Efficient Natural Language Processing}}. \bibinfo{pages}{83--88}.
\newblock


\bibitem[Zamani et~al\mbox{.}(2022)]%
        {zamani2022stochastic}
\bibfield{author}{\bibinfo{person}{Hamed Zamani}, \bibinfo{person}{Michael Bendersky}, \bibinfo{person}{Donald Metzler}, \bibinfo{person}{Honglei Zhuang}, {and} \bibinfo{person}{Xuanhui Wang}.} \bibinfo{year}{2022}\natexlab{}.
\newblock \showarticletitle{Stochastic Retrieval-Conditioned Reranking}. In \bibinfo{booktitle}{\emph{ICTIR}}. \bibinfo{pages}{81--91}.
\newblock


\bibitem[Zhang et~al\mbox{.}(2023)]%
        {zhang2023rank}
\bibfield{author}{\bibinfo{person}{Xinyu Zhang}, \bibinfo{person}{Sebastian Hofst{\"a}tter}, \bibinfo{person}{Patrick Lewis}, \bibinfo{person}{Raphael Tang}, {and} \bibinfo{person}{Jimmy Lin}.} \bibinfo{year}{2023}\natexlab{}.
\newblock \showarticletitle{Rank-without-GPT: Building GPT-Independent Listwise Rerankers on Open-Source Large Language Models}.
\newblock \bibinfo{journal}{\emph{arXiv preprint arXiv:2312.02969}} (\bibinfo{year}{2023}).
\newblock


\bibitem[Zhang et~al\mbox{.}(2021)]%
        {zhang2021learning}
\bibfield{author}{\bibinfo{person}{Yue Zhang}, \bibinfo{person}{ChengCheng Hu}, \bibinfo{person}{Yuqi Liu}, \bibinfo{person}{Hui Fang}, {and} \bibinfo{person}{Jimmy Lin}.} \bibinfo{year}{2021}\natexlab{}.
\newblock \showarticletitle{Learning to Rank in the Age of Muppets: Effectiveness–Efficiency Tradeoffs in Multi-Stage Ranking}. In \bibinfo{booktitle}{\emph{Proceedings of the Second Workshop on Simple and Efficient Natural Language Processing}}. \bibinfo{pages}{64--73}.
\newblock


\bibitem[Zhuang et~al\mbox{.}(2023b)]%
        {zhuang2023beyond}
\bibfield{author}{\bibinfo{person}{Honglei Zhuang}, \bibinfo{person}{Zhen Qin}, \bibinfo{person}{Kai Hui}, \bibinfo{person}{Junru Wu}, \bibinfo{person}{Le Yan}, \bibinfo{person}{Xuanhui Wang}, {and} \bibinfo{person}{Michael Berdersky}.} \bibinfo{year}{2023}\natexlab{b}.
\newblock \showarticletitle{Beyond Yes and No: Improving Zero-Shot LLM Rankers via Scoring Fine-Grained Relevance Labels}.
\newblock \bibinfo{journal}{\emph{arXiv preprint arXiv:2310.14122}} (\bibinfo{year}{2023}).
\newblock


\bibitem[Zhuang et~al\mbox{.}(2023a)]%
        {zhuang2023open}
\bibfield{author}{\bibinfo{person}{Shengyao Zhuang}, \bibinfo{person}{Bing Liu}, \bibinfo{person}{Bevan Koopman}, {and} \bibinfo{person}{Guido Zuccon}.} \bibinfo{year}{2023}\natexlab{a}.
\newblock \showarticletitle{Open-source Large Language Models are Strong Zero-shot Query Likelihood Models for Document Ranking}.
\newblock \bibinfo{journal}{\emph{arXiv preprint arXiv:2310.13243}} (\bibinfo{year}{2023}).
\newblock


\bibitem[Zhuang et~al\mbox{.}(2023c)]%
        {zhuang2023setwise}
\bibfield{author}{\bibinfo{person}{Shengyao Zhuang}, \bibinfo{person}{Honglei Zhuang}, \bibinfo{person}{Bevan Koopman}, {and} \bibinfo{person}{Guido Zuccon}.} \bibinfo{year}{2023}\natexlab{c}.
\newblock \showarticletitle{A Setwise Approach for Effective and Highly Efficient Zero-shot Ranking with Large Language Models}.
\newblock \bibinfo{journal}{\emph{arXiv preprint arXiv:2310.09497}} (\bibinfo{year}{2023}).
\newblock


\end{thebibliography}
